\documentclass[pra,superscriptaddress,amsmath,11pt]{revtex4-2}
\usepackage{graphicx}
\usepackage{subfigure}
\usepackage{bm}
\usepackage{bbm}
\usepackage{amssymb}

\usepackage{color}

\begin{document}
	
\title{Steered quantum coherence and quantum Fisher information in spin-chain system}
\author{Biao-Liang Ye}
\email{biaoliangye@gmail.com}
\affiliation{Quantum Information Research
Center, Shangrao Normal University, Shangrao 334001, China}

\author{Yao-Kun Wang}
\affiliation{College of Mathematics, Tonghua Normal University, Tonghua, Jilin 134001, China}

\author{Shao-Ming Fei}
\affiliation{School of Mathematical Sciences, Capital Normal University, Beijing 100048, China}
\affiliation{Max-Planck-Institute for Mathematics in the Sciences, 04103 Leipzig, Germany}


\begin{abstract}
In this paper, we investigate steered quantum coherence, i.e., the $l_1$ norm of steered coherence and the relative entropy of steered coherence,
and the quantum Fisher information in the Gibbs state of two-qubit $XXZ$ systems.
Their variations with respect to the temperature, external magnetic field, and interaction intensities are analyzed both analytically and numerically in detail.
The similar behaviors among these three quantum measures in the
$XXZ$ model are presented.
\medskip

\end{abstract}

\maketitle

\section{Introduction}
Spin chain models are of significance in such as quantum phase transitions (QPTs) in condensed matter systems \cite{Sachdev1999}.
Recently, it has attracted much attention to study spin chain systems in the framework of quantum information theories \cite{Modi2012,Adesso2016}.
The quantum information plays an important role in quantum communication and quantum computation \cite{Nielsen2010}. Especially, the quantum correlations like quantum entanglement and quantum discord \cite{Ollivier2001,Dakic2010,Luo2008,Rulli2011} play the key roles in quantum teleportation \cite{Bennett1993}, quantum secure direct communication \cite{Wang2005}, quantum network \cite{Yu2020}, remote state preparation \cite{Dakic2012}, quantum state discrimination \cite{Li2012} and quantum state merging \cite{Madhok2011}.

As one of the important measures of quantum entanglement, concurrence
has been used to investigate the $XY$ model \cite{Cakmak2015,Barouch1970,Barouch1971}.
The relationship between the critical point
and the entanglement in $XXZ$ chain \cite{AYurischev2020} has been also illustrated.
The first and second-order transition points have been displayed in terms of the two-spin entanglement \cite{Arnesen2001}.
Furthermore, quantum discord shows some advantages in
characterizing quantum critical points
even at higher temperature \cite{Maziero2010}.
In Ref. \cite{Ye2017} the authors studied
the quantum phase transitions in $XY$ chain based on
one-way quantum deficit.
A topological phase transition is revealed
by the geometric discord in Ref. \cite{Shan2014}.
The measurement-induced disturbance has been used to
show the first and second-order quantum critical systems \cite{Altintas2012}.
In ref. \cite{Sha2018}, $l_1$ norm coherence, skew information, and Bures distance of geometry discord (BGD) have been used to capture the quantum coherence and correlation of thermal states in the extended $XY$ spin chain. They have shown that the susceptibility of the skew information and BGD is a genuine indicator of quantum phase transitions in the model. However, the $l_1$ norm is trivial for the factorization. Li and Lin \cite{li2016quantum} have taken
local quantum coherence based on Wigner-Yanase skew information to investigate quantum phase transitions on the one-dimensional Hubbard, $XY$ spin chain with three-spin interaction and Su-Schrieffer-Heeger models. The measures are successful in detecting different types of QPTs in these spin and fermionic systems. Yin et. al. \cite{yin2018quantum}
used the relative entropy and quantum skew information for quantum coherence to study the dynamics of tripartite systems in a quantum-critical environment.
They shew the frozen for the initial $W$ state coupled to an $XY$ spin-chain environment as the system–environment coupling satisfies certain conditions.
Zhang and Xu \cite{zhang2017quantum} used a quantum renormalization group-based method
to $XY$ spin chain with Dzyaloshinskii-Moriya interaction
and show the relative entropy of coherence increases with Dzyaloshinskii-Moriya interaction strength at zero temperature, as well as at finite temperature.
In ref. \cite{jafari2020dynamics} the relative entropy, $l_1$ norm of coherence, Wigner-Yanase skew and quantum Fisher information had been used to study the dynamics of
the one-dimensional $XY$ spin chain in the presence of a time-dependent transverse magnetic field. They show that independent of the initial state of the system and while the relative entropy of coherence, the $l_1$ norm of coherence, and quantum Fisher information are incapable, surprisingly, the Wigner-Yanase skew information dynamic can truly spotlight the equilibrium critical point.
Qin's group \cite{qin2018dynamics}
use the quantum renormalization group theory to study $XY$ spin systems based on quantum coherence. They find that the quantum coherence obeys a conservation relation, i.e., the quantum coherence of a three-block state that is the sum of its reduced state coherence.
Lu and Wang \cite{Lu2021} incorporate Heisenberg’s uncertainty principle into quantum multiparameter estimation by giving a trade-off relation between the measurement inaccuracies for estimating different parameters.
Moreover, some important results related to the quantum coherence in spin-chain models have been obtained \cite{Hu2021,Zhao2022,Liu2023}.
Different quantum measures had also been discussed in application to many systems and possibility reliability in experiment \cite{Streltsov2017,Zheng2018,Smirne2020}.

As above mentioned, quantum coherence is
a key resource in quantum information processing \cite{Baumgratz2014,Karpat2014}.
It has been characterized by the quantum critical points in
$XY$ spin chains \cite{Li2020}. However, few measures are taken
to capture the quantum behaviors of the $XXZ$ system. It remains
key to show quantum behaviors via different quantum measures.
In this work, we devote to studying the $XXZ$ model based on the $l_1$ norm of steered coherence, the relative entropy of steered coherence \cite{Hu2020}, as well as the quantum Fisher information.
In Sec. II we briefly introduce the basic definitions
and notations. In Sec. III, the $XXZ$ model is introduced and 
the $l_1$ norm of steered coherence, the relative entropy of steered coherence, and quantum Fisher information of the Gibbs states of two-qubit XXZ systems are calculated and analyzed.
We conclude the paper in Sec. IV.

\section{Preliminary and Notations}
We first recall the basic concepts and
notations of steered quantum coherence and
quantum Fisher information.

i) Steered quantum coherence (SQC) \cite{Mondal2017}.
Let $\varrho_{AB}$ be a bipartite quantum qubit state shared by Alice and Bob,
and $\sigma^\mu$, $\mu=x, y, z$, the Pauli operators.
Alice carries out one of the pre-agreed measurements $\{\sigma^\mu\}_{\mu=x, y, z}$
on the qubit $A$ and communicates to Bob her choice $\sigma^\mu$. After the measurement, Bob's system collapses to the
ensemble states $\{p_{\mu, a}, \varrho_{B|\Pi_\mu^a}\}$, where $p_{\mu, a}
=\textrm{Tr}(\Pi_\mu^a \varrho_{AB})$ is the probability of Alice's measurement outcome
$a\in\{0,1\}$, $\varrho_{B|\Pi_\mu^a}= \textrm{Tr}_A (\Pi_\mu^a\varrho_{AB})
/p_{\mu, a}$ is the Bob's conditional state, and $\Pi_\mu^{a}=
[I_2+(-1)^a \sigma^\mu]/2$ is the measurement operator with
$I_2$ the $2\times 2$ identity operator.

Associated to Alice's observable $\sigma^\mu$, Bob may measure the
coherence of the ensemble $\{p_{\mu, a}, \varrho_{B|\Pi_\mu^a}\}$ with
respect to the eigenbasis of either one of the remaining two Pauli
operators. After Alice's all possible measurements
$\{\Pi_\mu^a\}_{\mu=x, y, z}$ with equal probability, the SQC at Bob's
hand can be defined as the following averaged quantum coherence,
\begin{equation} \label{eq2a-2}
 C^{na}(\varrho_{AB})= \frac{1}{2}\sum_{\mu, \nu, a\atop \mu\neq\nu}
                    p_{\mu, a} C^{\sigma^\nu}(\varrho_{B|\Pi_\mu^a}),
\end{equation}
where $C^{\sigma^\nu}(\varrho_{B|\Pi_\mu^a})$ is the coherence of
$\varrho_{B|\Pi_\mu^a}$ defined in the reference basis spanned by the
eigenbases of $\sigma^\nu$.

In this paper, we take the $l_1$ norm of steered coherence, and
the relative entropy of steered coherence as the measures of steered coherence.
By denoting $\{|\psi_i\rangle\}$ the eigenbases of
$\sigma^\nu$, $l_1$ norm of coherence and the relative
entropy of coherence are given by \cite{Hu2020}
\begin{eqnarray} \label{eq2a-3}
  && C_{l_1}^{\sigma^\nu}(\varrho)= \sum_{i\neq j}|\langle\psi_i|\varrho|\psi_j\rangle|,\\
  && C_{re}^{\sigma^\nu}(\varrho)= -\sum_i \langle\psi_i|\varrho|\psi_i\rangle
                               \log\langle\psi_i|\varrho|\psi_i\rangle-S(\varrho),
\end{eqnarray}
respectively, where $S(\varrho)=-\textrm{Tr} (\varrho\log \varrho)$ is the von Neumann
entropy. Without loss of generality, the $\log$ is in base 2. The corresponding SQCs
are then given by (\ref{eq2a-2}).

ii) Quantum Fisher information (QFI) \cite{Ye2018b}.
In general phase estimation scenarios, the evolution of a quantum state $\varrho$ under a unitary transformation
is described as
$\varrho_\theta=e^{-i A\theta}\varrho e^{i A\theta}$,
where $\theta$ is the phase shift and $A$ is
an operator. The estimation accuracy for $\theta$
is limited by the quantum
Cram\'{e}r-Rao inequality:
\begin{equation}
	\Delta\hat{\theta}\ge\frac{1}{\sqrt{\nu \mathcal{F}(\varrho_\theta)}},
\end{equation}
where $\hat{\theta}$ denotes the unbiased
estimator for $\theta$, $\nu$ is the number of times the measurement repeated,
and $\mathcal{F}(\varrho_\theta)$ is the so-called
QFI. The QFI of a state $\varrho$ with respect to an observable $\hat{O}$ is defined by
\begin{equation}\label{f}
\mathcal{F}(\varrho, \hat{O})=2\sum_{m,n}\frac{(p_m-p_n)^2}{(p_m+p_n)}|\langle m|\hat{O}|n\rangle|^2,
\end{equation}
where $p_m (p_n)$ and $|m\rangle$ and $(|n\rangle)$ are
the eigenvalues and eigenvectors, respectively, of the
density matrix $\varrho$ which is used as a probe state to estimate $\theta$.

\section{$XXZ$ spin chain}

We consider the Heisenberg $XXZ$ two-qubit anisotropic spin chain
with Hamiltonian \cite{AYurischev2020},
\begin{eqnarray}
	H=-\frac12[J(\sigma_x^A\sigma_x^B+
	\sigma_y^A\sigma_y^B)+J_z\sigma_z^A\sigma_z^B]
	-\frac12B(\sigma_z^A+\sigma_z^B),
\end{eqnarray}
where $B$ is the uniform external magnetic field, $J$ and $J_z$ are the coupling parameters,
$\sigma_{x}^{A,B}, \sigma_y^{A,B}$ and $\sigma_z^{A,B}$ the Pauli matrices
associated with qubits $A$ and $B$, respectively.
By calculating the eigenvalues $E_i$ of $H$, $i=1,...,4$, one gets partition function $Z=\sum_i\exp(-E_i/T)$ \cite{AYurischev2020},
\begin{equation} \label{part}
	Z=2(\exp({J_z/2T})\cosh\frac{B}{T}+\exp({-J_z/2T})\cosh\frac{J}{T}).
\end{equation}
The Gibbs density matrix is given by
\begin{eqnarray}
	\varrho_{AB}=e^{-H/T}/Z=\left(
	\begin{array}{cccc}
		a & 0 & 0 & 0\cr
		0 & b & v & 0\cr
		0 & v & b & 0\cr
		0 & 0 & 0 & d\cr
	\end{array}\right),
\end{eqnarray}
where $T$ is the temperature,
\begin{eqnarray} \label{x}
	a&=&\frac{1}{Z}\exp{[(J_z/2+B)/T]},
	b = \frac{1}{Z}\exp[{-J_z/2T]}\cosh(J/T),\cr
	d&=&\frac{1}{Z}\exp{[(J_z/2-B)/T]},
	v = \frac{1}{Z}\exp[{-J_z/2T]}\sinh(|J|/T).\nonumber
\end{eqnarray}

According to the definitions of SQC and QFI, we have the following results of the state $\varrho_{AB}$.

(i) The $l_1$ norm of steered coherence (SCn) is given by
\begin{equation}\label{scn}
SCn=\sqrt{(a-d)^2+4 v^2}+|a-b|+|b-d|+2|v|.
\end{equation}

(ii) The relative entropy of steered coherence (SCRE) is given by
\begin{eqnarray}
	&&SCRE=
	\frac14[(1-a-2b+3d)\log(1-a-2b+3d)\nonumber\\
	&&+(1+3a-2b-d)\log(1+3a-2b-d)]\nonumber\\
	&&+\frac12(1-a+2b-d)\log(1-a+2b-d)\nonumber\\
	&&+\sum_{\pm}(1\pm\sqrt{(a-d)^2+4 v^2})\log(1\pm\sqrt{(a-d)^2+4 v^2}).\nonumber\\
\end{eqnarray}

(iii) The quantum Fisher information (QFI) is given by
\begin{eqnarray}
	QFI&&=\frac{m}{n},
\end{eqnarray}
where
\begin{eqnarray}
	m&&=e^{\frac{2 J_z}{T}} (\sinh \frac{| J| }{T}+\cosh \frac{J}{T})(e^{-\frac{B}{T}}-4 e^{\frac{ B}{T}}+e^{\frac{3 B}{T}})\nonumber\\
	&&-e^{\frac{J_z}{T}}(e^{\frac{2 B}{T}}+1)  (\sinh \frac{| J| }{T}+\cosh \frac{J}{T})^2\nonumber\\
	&&+2 e^{\frac{2 B}{T}} (\sinh \frac{| J| }{T}+\cosh \frac{J}{T})^3\nonumber\\
	&&+e^{\frac{3 J_z}{T}}(e^{\frac{2 B}{T}}+1),\nonumber\\[1mm]
	n&&=(e^{J_z/T} \cosh \frac{B}{T}+\cosh \frac{J}{T}) (\sinh \frac{| J| }{T}+\cosh \frac{J}{T}\nonumber\\
	&&+e^{\frac{B+J_z}{T}}) [e^{B/T} (\sinh \frac{| J| }{T}+\cosh \frac{J}{T})+e^{J_z/T}].\nonumber
\end{eqnarray}

\begin{figure}[htbp!]
\centering
\subfigure[$B=1$]{
\includegraphics[width=.23\textwidth]{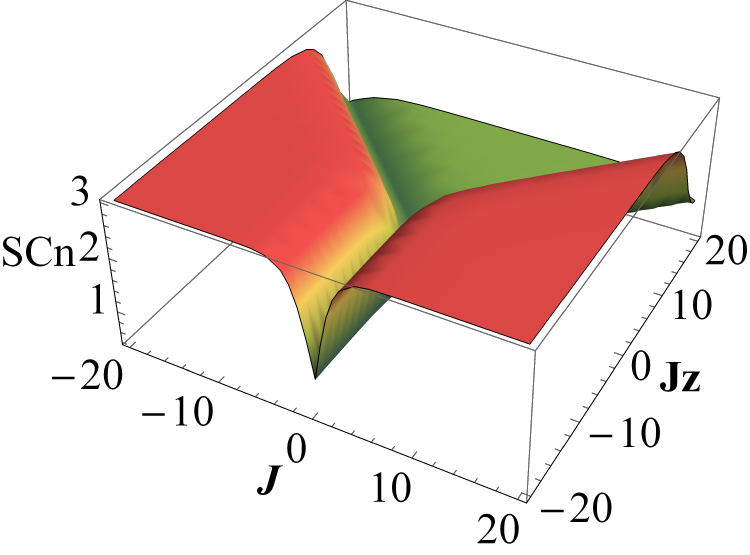}
}
\subfigure[$B=2$]{
\includegraphics[width=.23\textwidth]{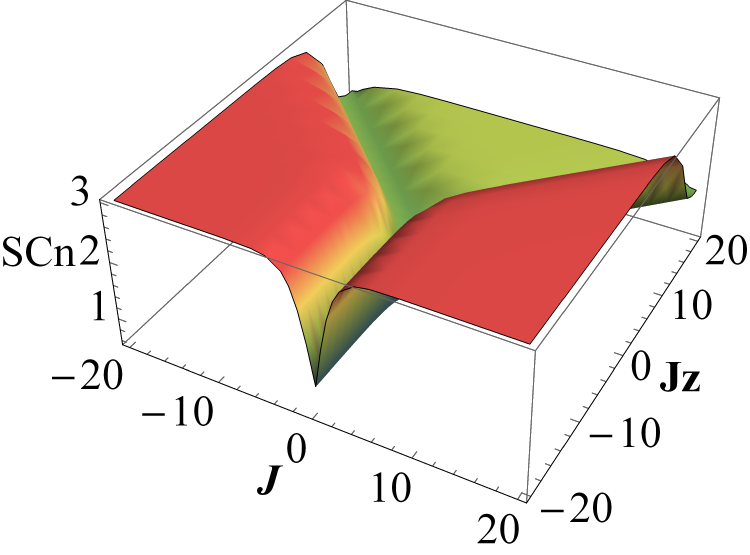}
}
\subfigure[$B=3$]{
\includegraphics[width=.23\textwidth]{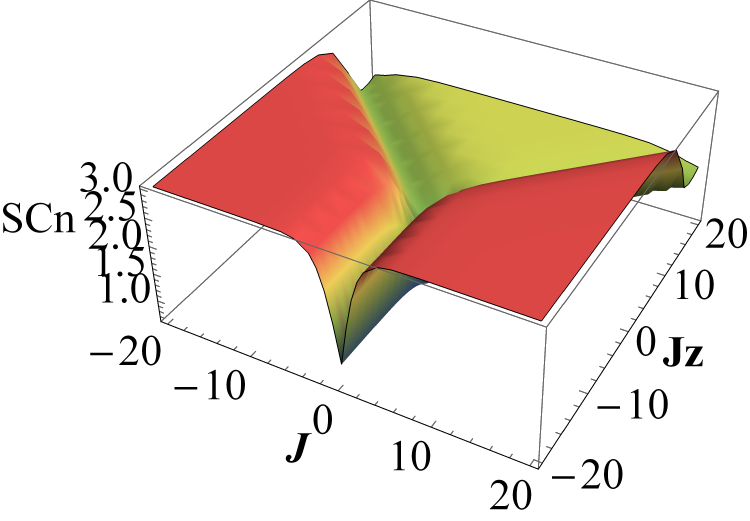}
}

\subfigure[$B=5$]{
\includegraphics[width=.23\textwidth]{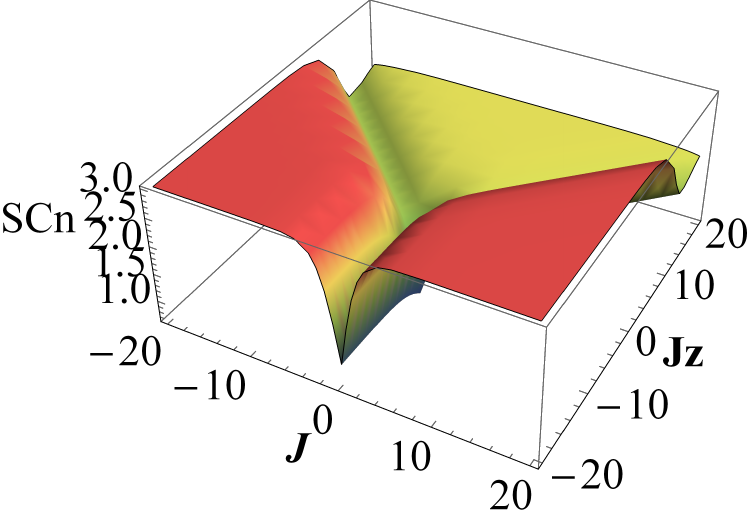}
}
\subfigure[$B=8$]{
\includegraphics[width=.23\textwidth]{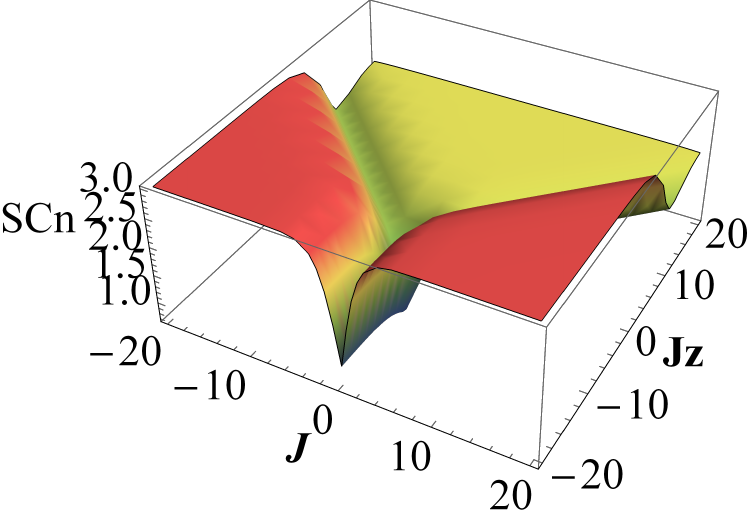}
}
\subfigure[$B=10$]{
\includegraphics[width=.23\textwidth]{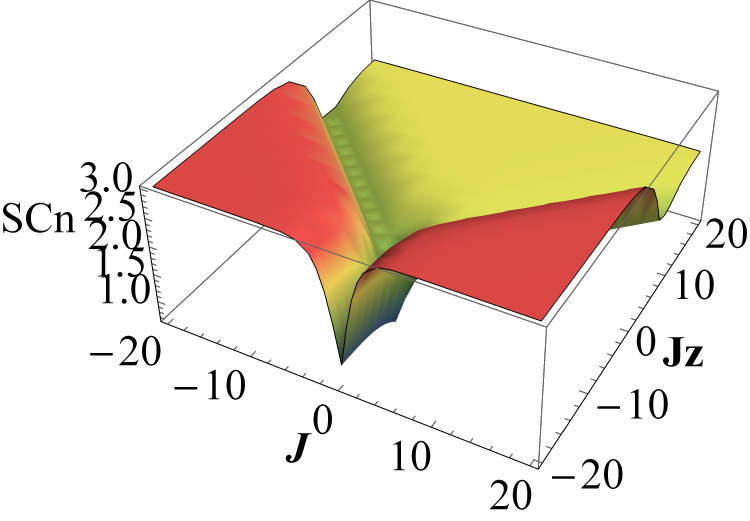}
}
\caption{The $l_1$ norm of steered coherence vs the $J$ and $J_z$ for
$T=2$ and $B=1, 2, 3, 5, 8$, and $10$ as shown above.}
\label{fig1}
\end{figure}

We illustrate the $l_1$ norm of steered coherence
with respect to $J$ and $J_z$ for $T=2$, $B=1, 2, 3, 5, 8$ and $10$, respectively, in Fig. \ref{fig1}.
Obviously, the SCn is symmetric around $J=0$. One can see that the SCn arrives at the maximum of 3.
As $B$ increases, the triangle zone expands and the SCn increases at the same time.

\begin{figure}[htbp!]
\centering
\subfigure[$B=1$]{
\includegraphics[width=.23\textwidth]{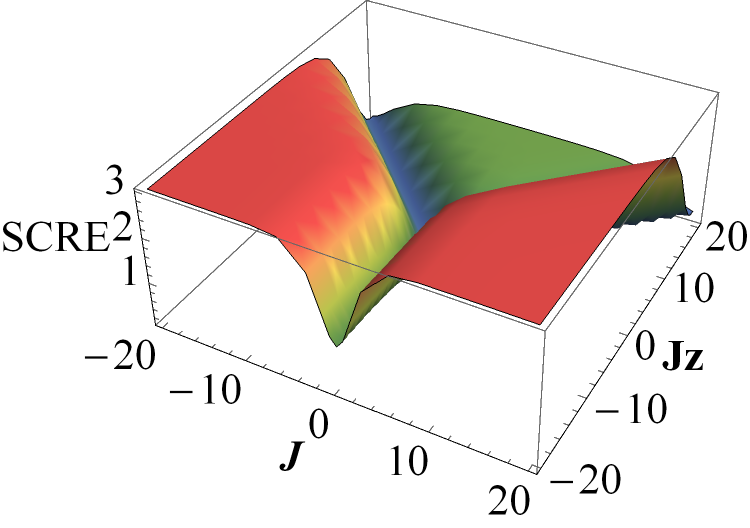}
}
\subfigure[$B=2$]{
\includegraphics[width=.23\textwidth]{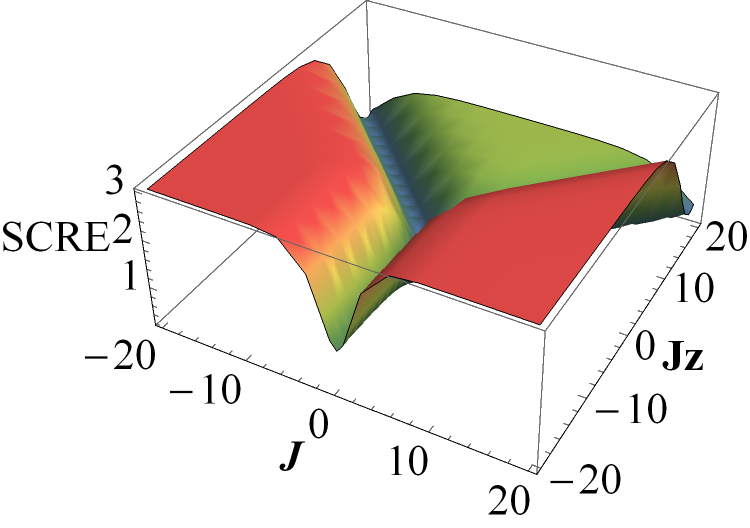}
}
\subfigure[$B=3$]{
\includegraphics[width=.23\textwidth]{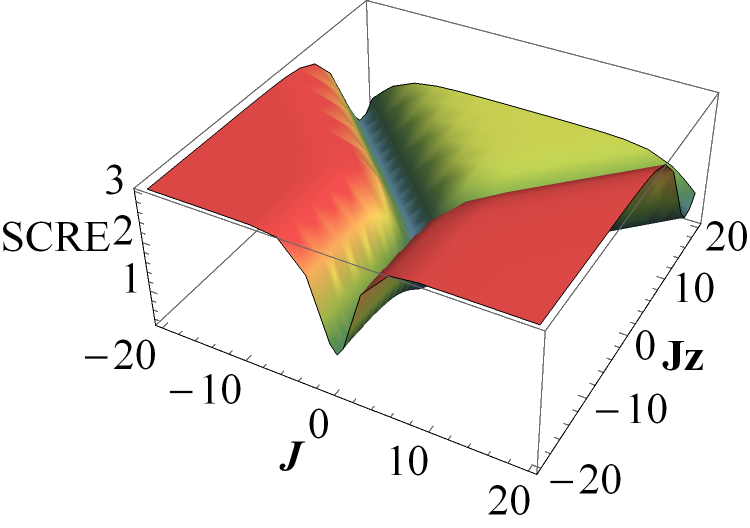}
}

\subfigure[$B=5$]{
\includegraphics[width=.23\textwidth]{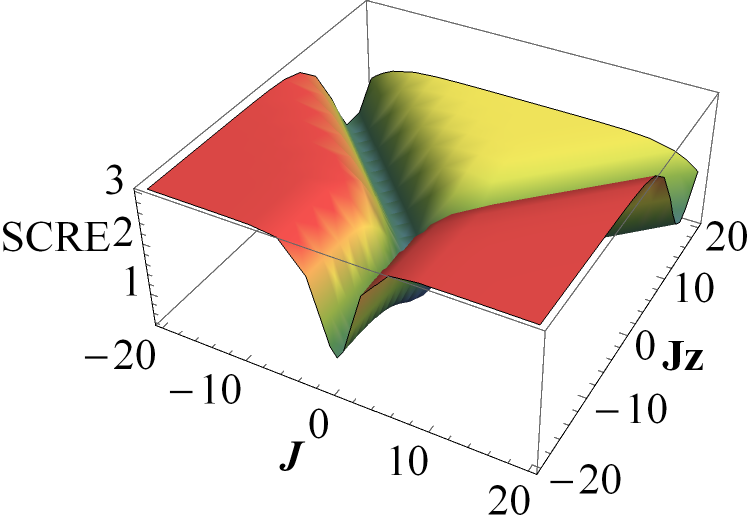}
}
\subfigure[$B=8$]{
\includegraphics[width=.23\textwidth]{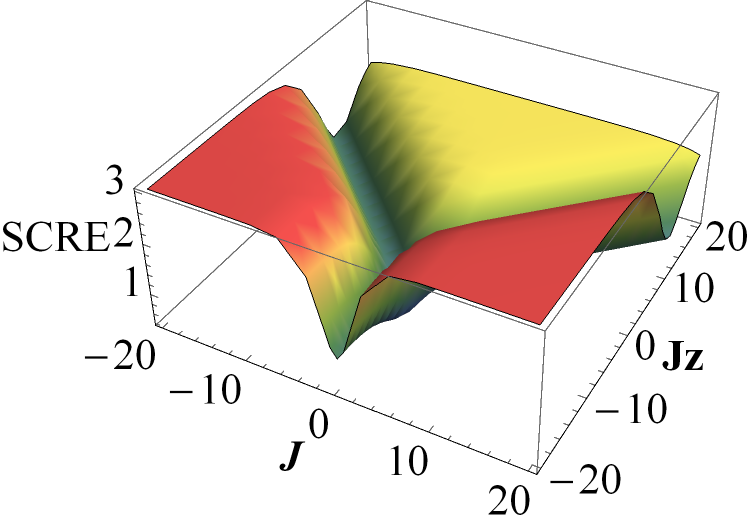}
}
\subfigure[$B=10$]{
\includegraphics[width=.23\textwidth]{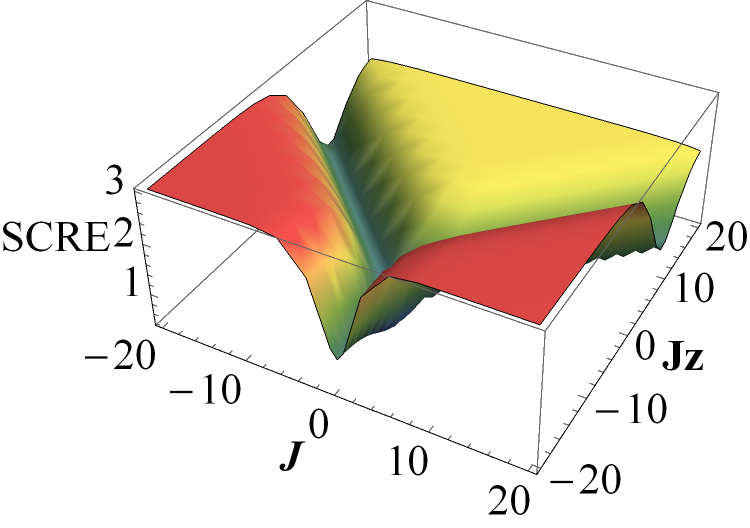}
}
\caption{The relative entropy of steered coherence vs the $J$ and $J_z$ for
$T=2$ and $B=1, 2, 3, 5, 8$, and $10$ as shown above.}
\label{fig2}
\end{figure}

The relative entropy of steered coherence
is shown in Fig. \ref{fig2}. One can see that the maximal value of 3
is attained. It is symmetric with respect to $J=0$.
As $B$ increases, the triangle zone becomes large.

\begin{figure}[htbp!]
\centering
\subfigure[$B=1$]{
\includegraphics[width=.23\textwidth]{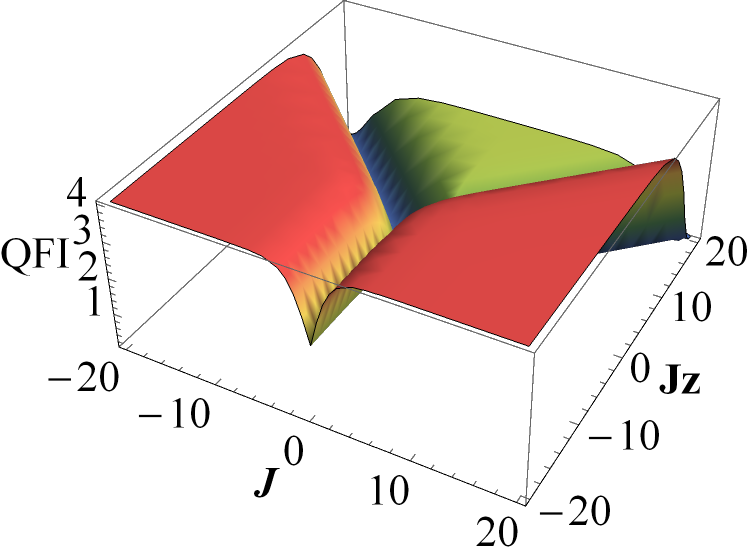}
}
\subfigure[$B=2$]{
\includegraphics[width=.23\textwidth]{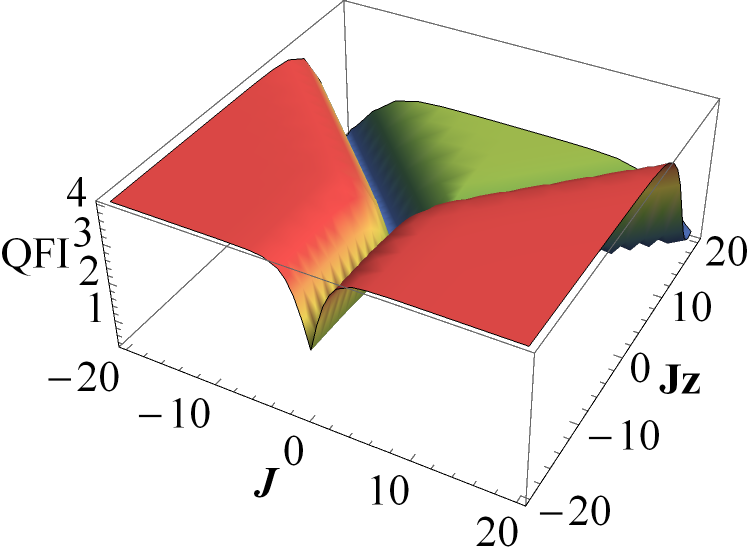}
}
\subfigure[$B=3$]{
\includegraphics[width=.23\textwidth]{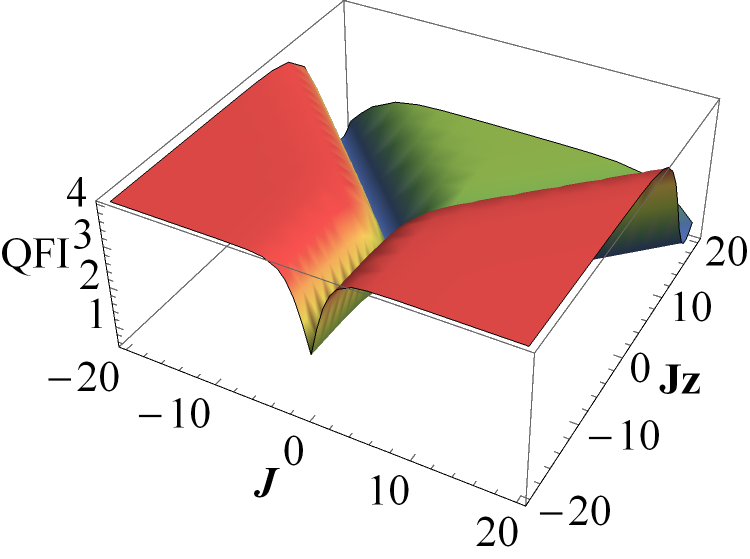}
}

\subfigure[$B=5$]{
\includegraphics[width=.23\textwidth]{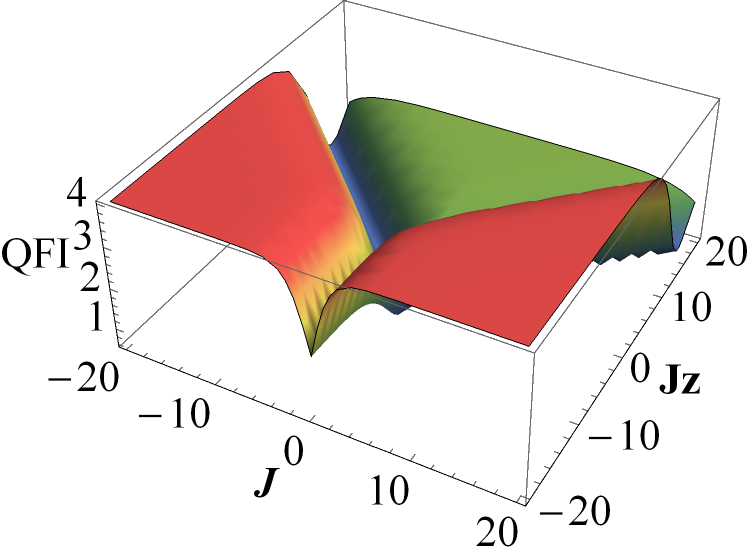}
}
\subfigure[$B=8$]{
\includegraphics[width=.23\textwidth]{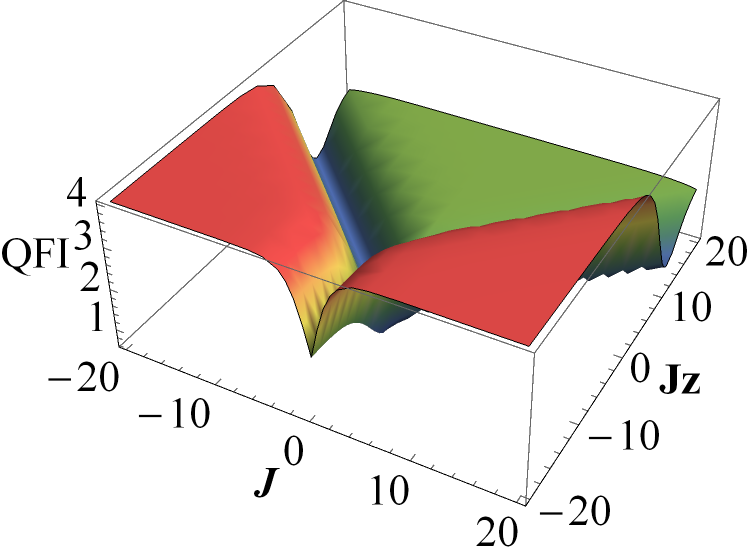}
}
\subfigure[$B=10$]{
\includegraphics[width=.23\textwidth]{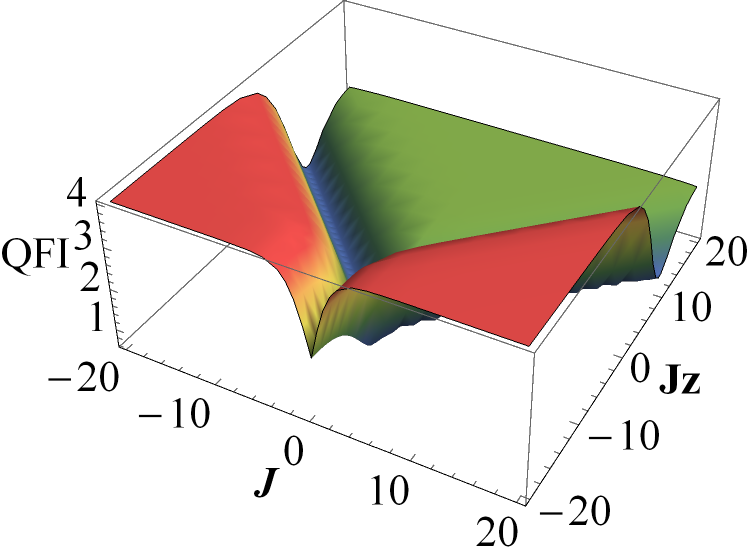}
}
\caption{Quantum Fisher information (QFI) against the $J$ and $J_z$ for
$T=2$ and $B=1, 2, 3, 5, 8$, and $10$ as shown above.}
\label{fig3}
\end{figure}

The quantum Fisher information for the Gibbs state $\varrho_{AB}$ is shown in Fig. \ref{fig3}.
The QFI is divided into three areas. The maximum QFI reaches 4. Similar
to the $l_1$ norm of steered coherence
and relative entropy of steered coherence, the triangle
area enlarges as the external magnetic field increases.
One sees the gullies on the boundaries of QFI are steep and sharp.

In Fig. 4 we show the
$l_1$ norm coherence of steered coherence against $B$ and $T$ for $J=10$
and $J_z=2,5,8$, respectively.
One can see that the SCn does not increase when the
$T$ increases if magnetic field $B$ is fixed.
Interestingly, for magnetic $B=0$ and $T\rightarrow0$,
we have the Bell state $(|01\rangle+|10\rangle)/\sqrt{2}$, and the SCn reaches
the maximal value 3. When $T$ and $B$ increase, the system
becomes mixed and the SCn decreases.
Moreover, around maximal value 3 the SCn decreases as $J_z$ increases from $2$ to $8$.
\begin{figure}[htbp!]
\centering
\subfigure[]{
\includegraphics[width=.23\textwidth]{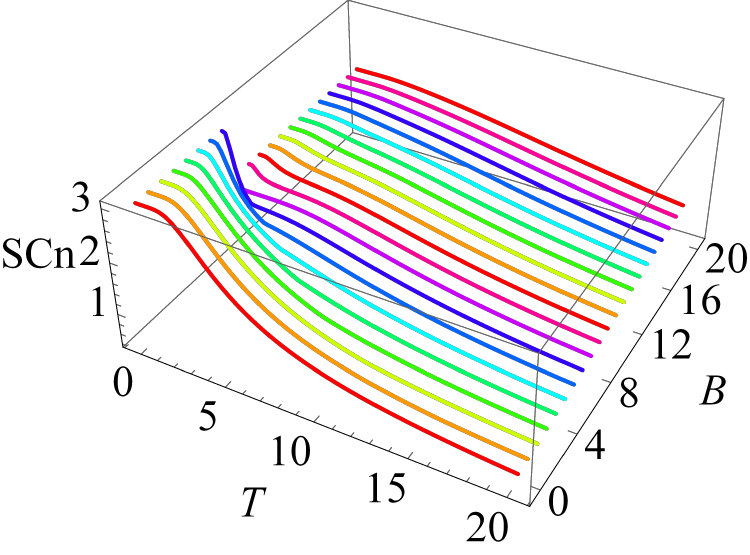}
}
\subfigure[]{
\includegraphics[width=.23\textwidth]{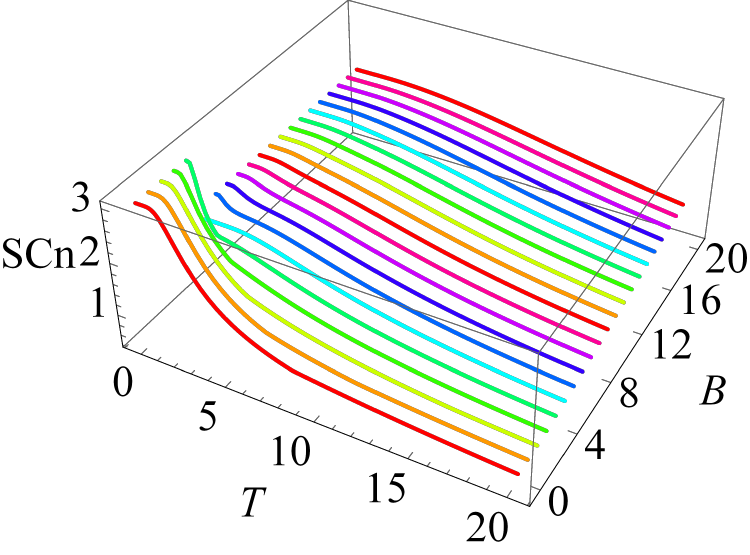}
}
\subfigure[]{
\includegraphics[width=.23\textwidth]{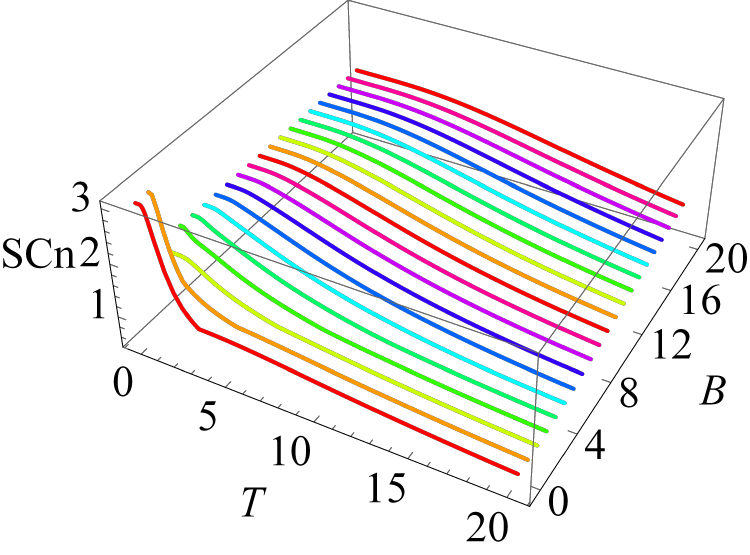}
}
\caption{The $l_1$ norm coherence of steered coherence against the $B$ and
$T$ for $J=10$ and $J_z=2, 5, 8$, respectively, from left to right.}
\label{line1}
\end{figure}

In Fig. 5, we plot the relative entropy of steered coherence against the $B$ and
$T$ for $J=10$ and $J_z=2, 5, 8,$ respectively.
One can see that the SCRE decreases quickly for low temperature $T$ and low magnetic $B$.
Similar to the SCn in Fig.4, for magnetic $B=0$ and $T\rightarrow0$, we have one Bell state and
the SCRE gets the value 3. Around the value 3 the SCRE decreases as $J_z$
increases from $2$ to $8$.
\begin{figure}[t]
\centering
\subfigure[]{
\includegraphics[width=.23\textwidth]{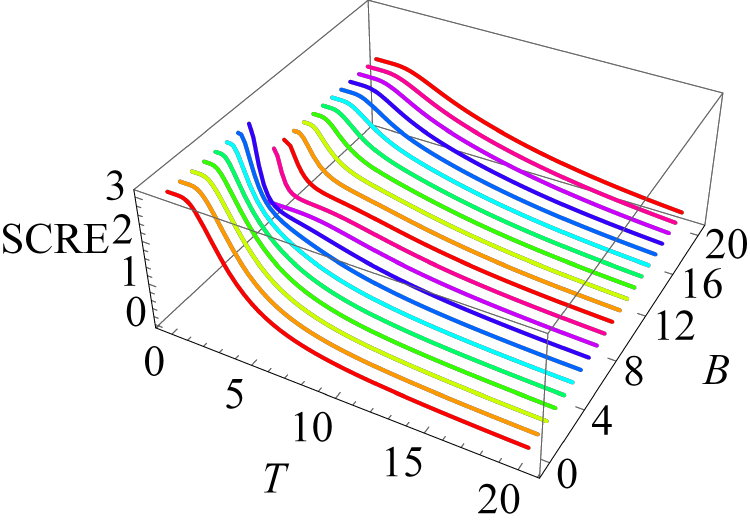}
}
\subfigure[]{
\includegraphics[width=.23\textwidth]{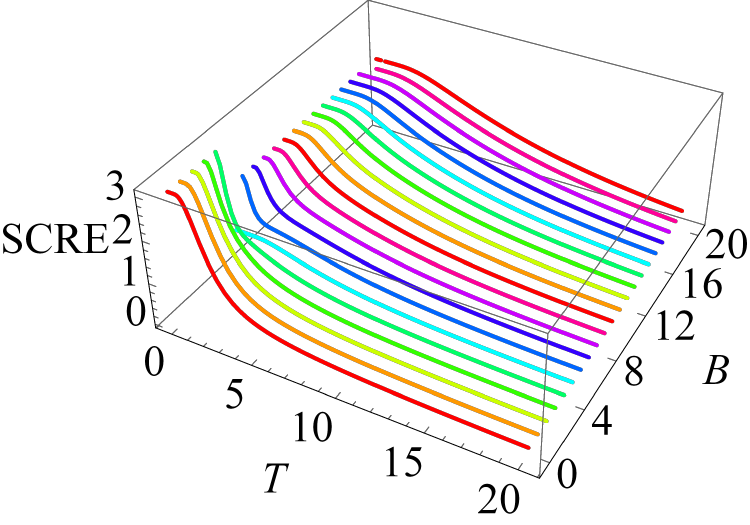}
}
\subfigure[]{
\includegraphics[width=.23\textwidth]{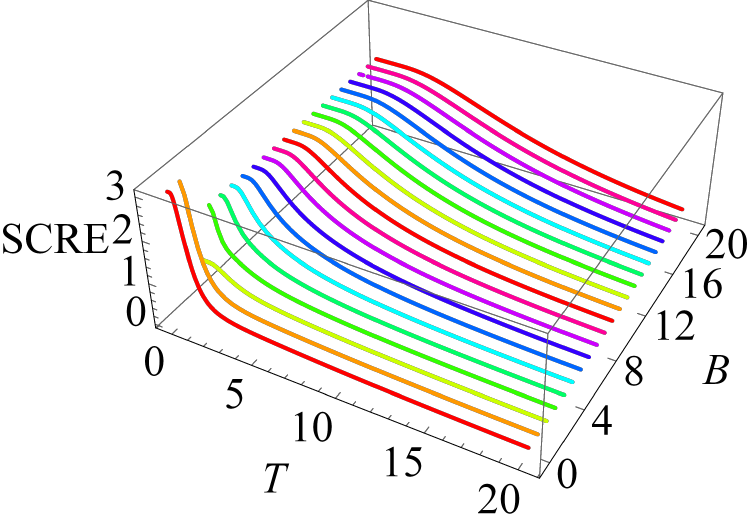}
}
\caption{The relative entropy of steered coherence against the $B$ and
$T$ for $J=10$ and $J_z=2, 5, 8,$ respectively, from left to right.}
\label{line2}
\end{figure}

Fig. 6 shows that the quantum Fisher information against the magnetic $B$ and
$T$ for $J=10$ and $J_z=2, 5, 8,$ respectively.
The behavior of QFI is similar to the ones of SCn and SCRE in Fig.4 and Fig.5, respectively.
For magnetic $B=0$ and $T\rightarrow0$, the system becomes a Bell state
and QFI attains its maximal value 4. Around this value $4$ the QFI decreases as $J_z$
increases from $2$ to $8$.
\begin{figure}[htbp!]
\centering
\subfigure[]{
\includegraphics[width=.23\textwidth]{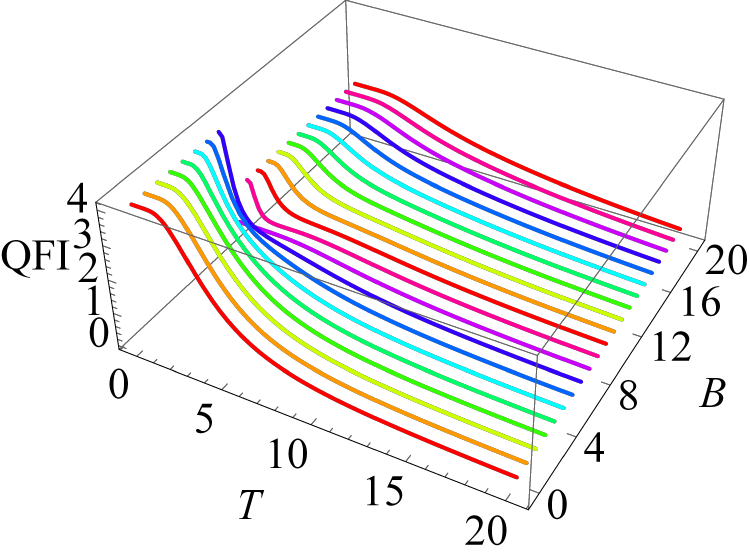}
}
\subfigure[]{
\includegraphics[width=.23\textwidth]{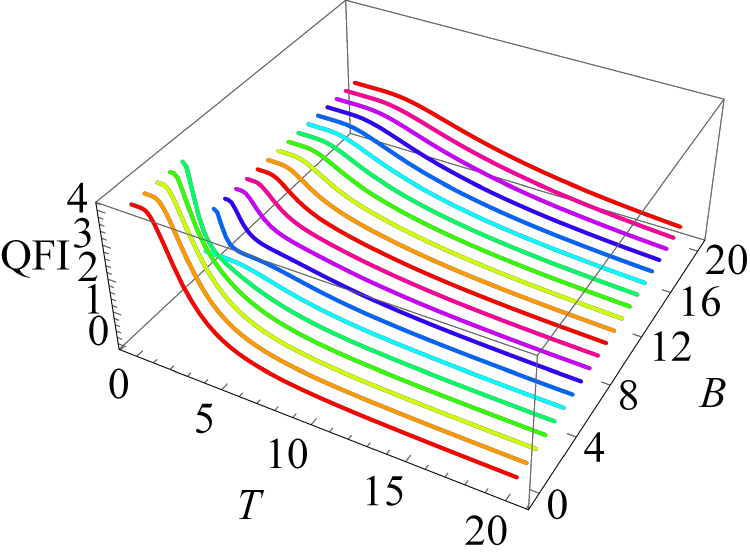}
}
\subfigure[]{
\includegraphics[width=.23\textwidth]{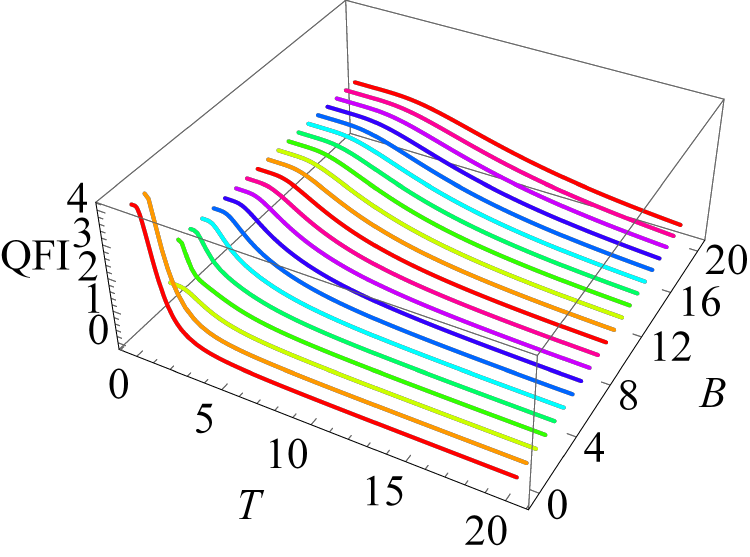}
}
\caption{Quantum Fisher information against the $B$ and
$T$ for $J=10$ and $J_z=2, 5, 8,$ respectively, from left to right.}
\label{line3}
\end{figure}

\begin{figure}[t]
\centering
\subfigure[]{
\includegraphics[width=.23\textwidth]{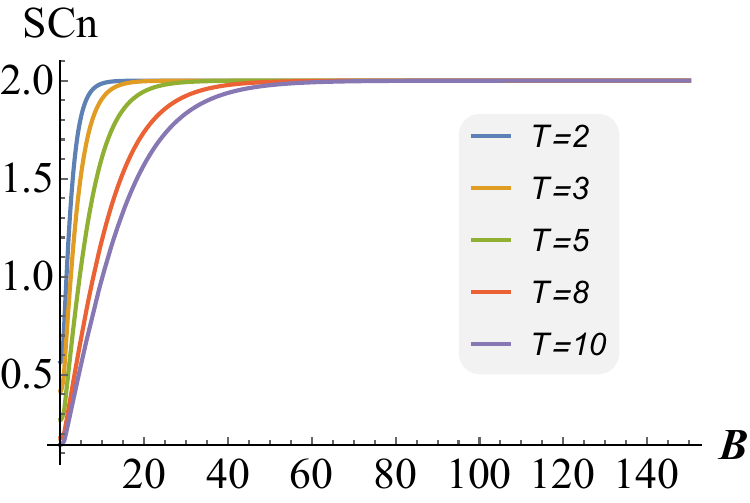}
}
\subfigure[]{
\includegraphics[width=.23\textwidth]{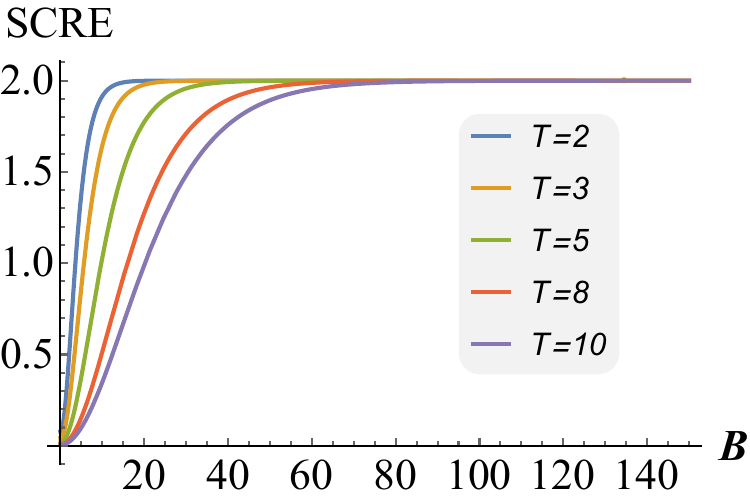}
}
\subfigure[]{
\includegraphics[width=.23\textwidth]{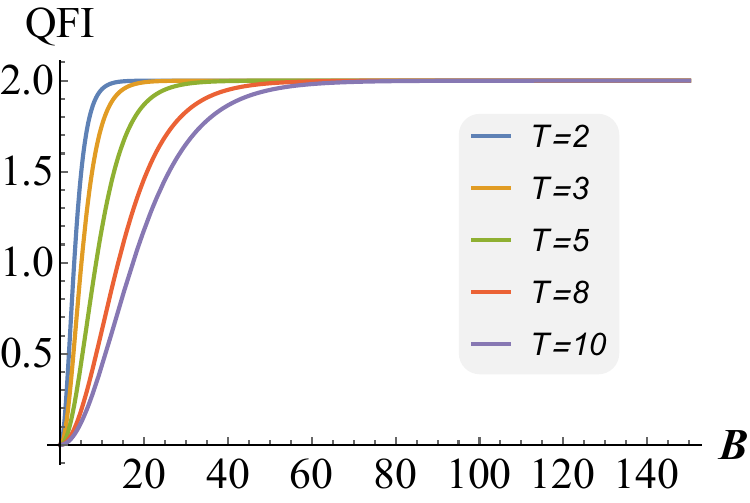}
}

\subfigure[]{
\includegraphics[width=.23\textwidth]{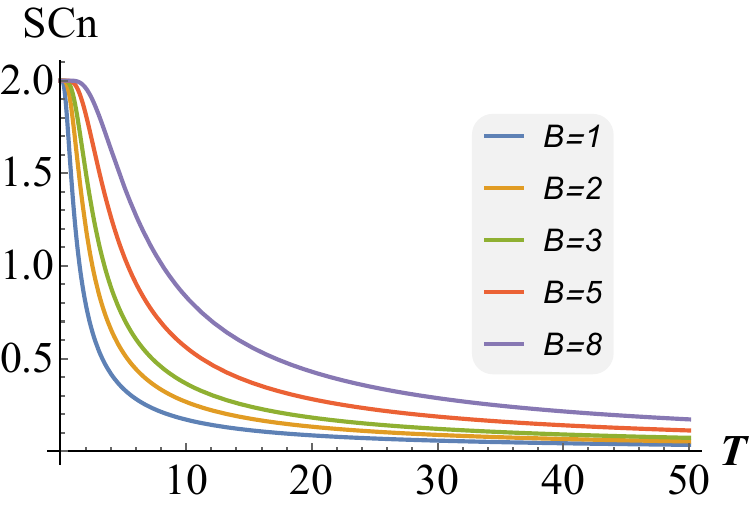}
}
\subfigure[]{
\includegraphics[width=.23\textwidth]{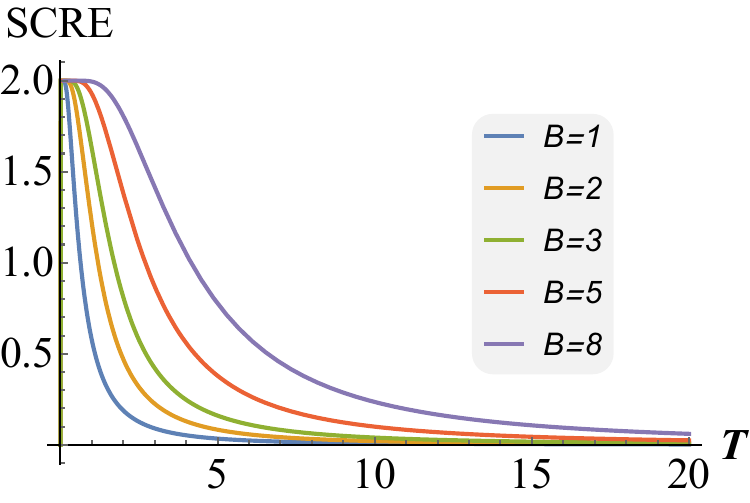}
}
\subfigure[]{
\includegraphics[width=.23\textwidth]{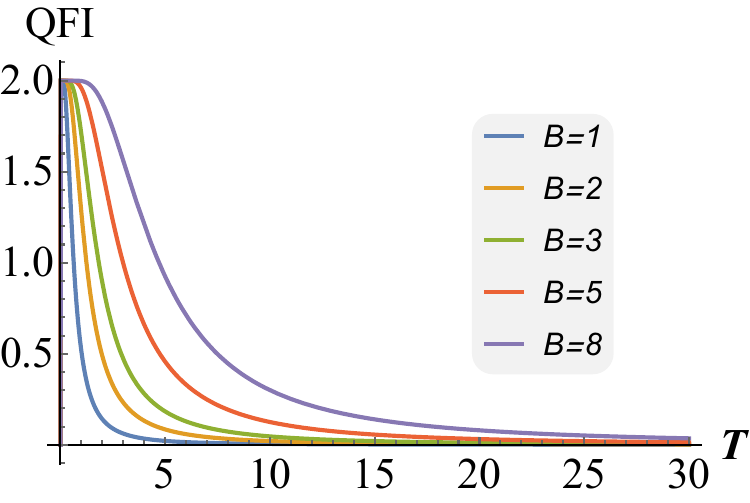}
}
\caption{The $l_1$ norm coherence of steered coherence, relative entropy
of steered coherence,
and quantum Fisher information vs $B$ for
$J=J_z=1$ and $T=2, 3, 5, 8,$ and $10$, Figs. (a)-(c).
The $l_1$ norm coherence of steered coherence, relative entropy
of steered coherence,
and quantum Fisher information vs $T$ for
$J=J_z=1$ and $B=1, 2, 3, 5, 8$, Figs. (d)-(f).}
\label{fig7}
\end{figure}

In Figure \ref{fig7}, we plot the variations of SCn, SCRE, and QFI with respect to magnetic field $B$ and temperature $T$, respectively. It can be seen that when the temperature is relatively low, the three measures all show the same tendency, rapid increasing to the maximum value and then keeping unchanged as the magnetic field increases. As the temperature increases, the slopes for the three measures reduce. These three measures SCn, SCRE, and QFI show similar behaviors. However, they are slightly different from SCn (a) in the case of low magnetic field strength. The slope is relatively large, while the slopes of (b) and (C) are small at the beginning and then increase rapidly.
At the same time, we consider the variation trend of these three measures with respect to temperature under the condition of $B=1,2,3,5,8$, represented by blue, orange, green, red, and purple, respectively. It is obvious that they decline rapidly at the beginning with the increase of temperature, and then gently approach a small value. All three measures start with a maximum value of 2, and as the temperature increases, they approach zero.

\begin{figure}[t]
\centering
\subfigure[]{
\includegraphics[width=.23\textwidth]{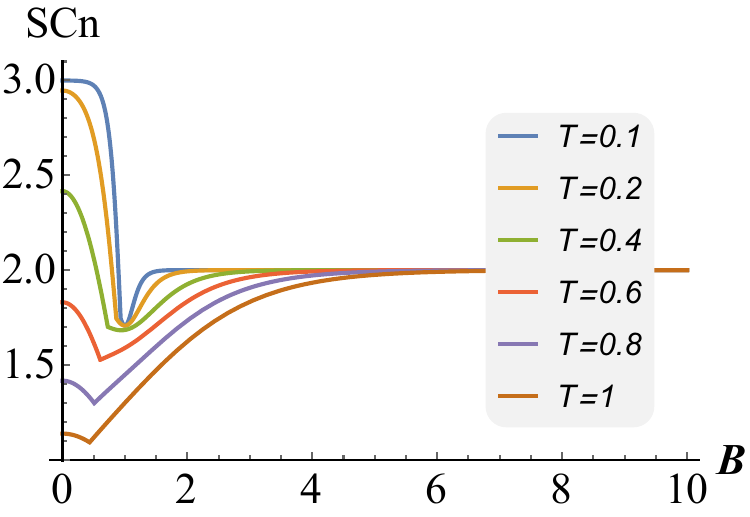}
}
\subfigure[]{
\includegraphics[width=.23\textwidth]{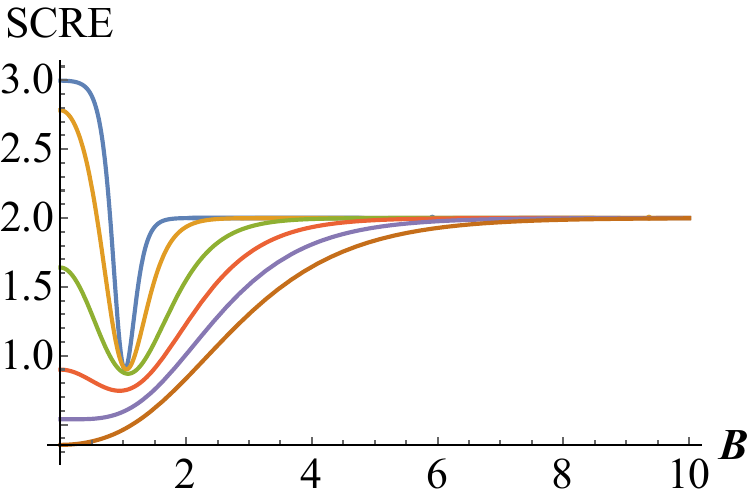}
}
\subfigure[]{
\includegraphics[width=.23\textwidth]{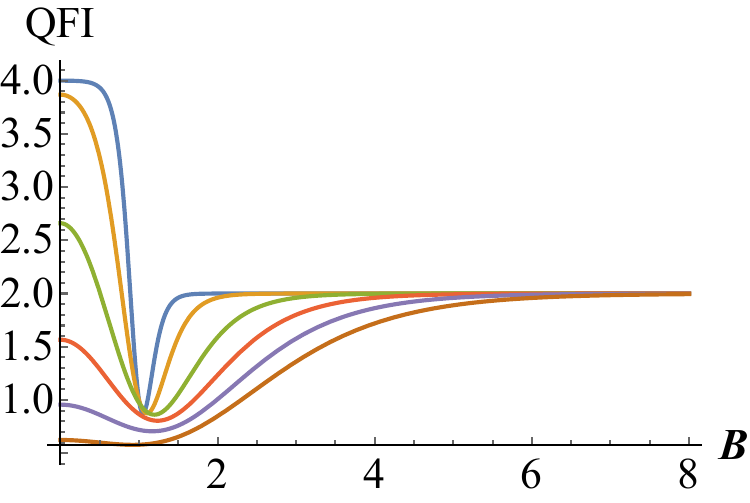}
}
\caption{The $l_1$ norm coherence of steered coherence, relative entropy
of steered coherence,
and quantum Fisher information vs $B$ for
$J=1, J_z=0$ and $T=0.1, 0.2, 0.4, 0.6, 0.8,$ and $1$.}
\label{fig8}
\end{figure}

In FIG. \ref{fig8}, we consider the behaviors of SCn, SCRE, and QFI under the variation of the magnetic field $B$ with $J=1$, $J_z=0$, and $T=0.1,0.2,0.4,0.6,0.8,1$ represented in blue, yellow, green, red, purple, and brown, respectively. It can be observed from the figures that SCn first reaches stability with the increase of $B$ when $T=0.1$, and then drops and rises rapidly, showing a small groove, and then reaches a maximum platform again. After the temperature reaches a certain limit, the small groove gradually disappears for larger $T$. One observes that with the increase of the magnetic field from the custom in the position of the magnetic field
$B = 1$ the SCRE shows a rapid drop and rises to value 2 for small $T$. However, for larger $T$ it also shows another behavior, it only gradually increases to the maximum. The QFI shows similar behaviors to SCRE, except that first it can reach the maximum value of 4, and for higher temperatures, it still reduces, then increases and reaches a stable value 2. All of SCn, SCRE, and QFI reach value of 2 for larger magnetic field $B$.

\begin{figure}[t]
\centering
\subfigure[]{
\includegraphics[width=.23\textwidth]{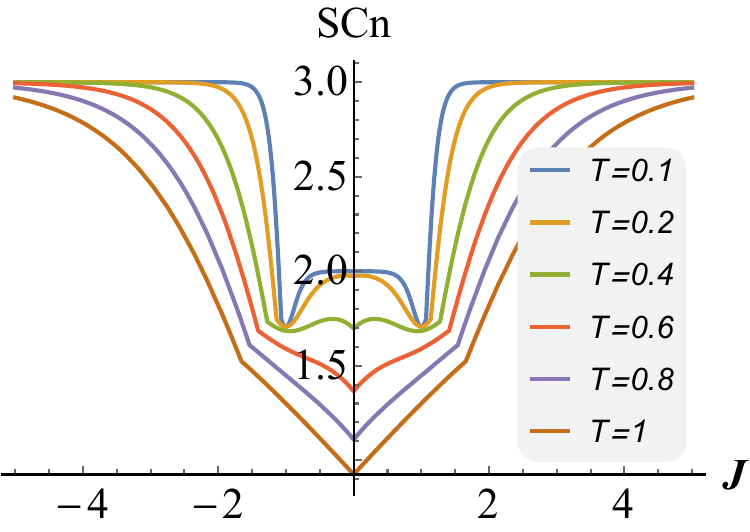}
}
\subfigure[]{
\includegraphics[width=.23\textwidth]{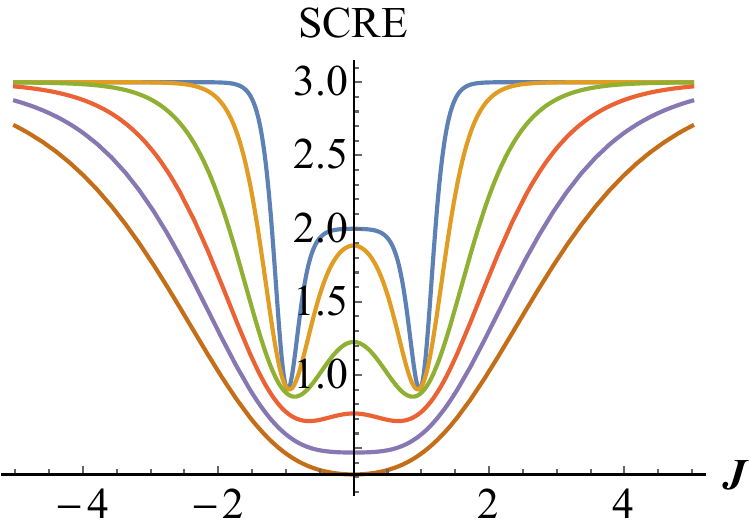}
}
\subfigure[]{
\includegraphics[width=.23\textwidth]{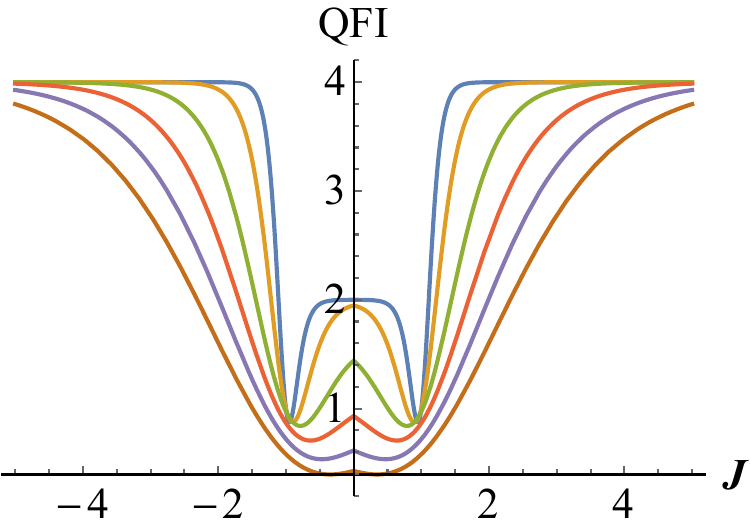}
}
\caption{The $l_1$ norm coherence of steered coherence, relative entropy
of steered coherence,
and quantum Fisher information vs the $J$ for
$B=1$, $J_z=0$ and $T=0.1, 0.2, 0.4, 0.6, 0.8,$ and $1$.}
\label{fig9}
\end{figure}

In Fig. \ref{fig9}, we plot the variations of the three measures with respect to the intensity $J$ at different temperatures $T=0.1, 0.2, 0.4, 0.6, 0.8, 1$, respectively, for fixed $B=1$ and $J_z=0$. One can see that the three measures show symmetric behaviors about $J$. In Fig.\ref{fig9} (a), SCn has grooves at 1 and -1, and has a maximum value of 2 between these two grooves. As the temperature increases, the value at $J=0$ decreases. In particular, two peaks appear at $[-1, 1]$ for $T=0.4$, which is symmetry about the vertical axis of $J=0$. Fig. \ref{fig9} (b) and (c) show similar behaviors, dropping from the maximum value, then rising and then falling again, finally reaching 3.
The QFI also gives this trend, except that the maximum value is 4.
Fig. \ref{fig9} (b) and (c) are slightly different, and the sharp point is taken at the position about the symmetry vertical axis of $J=0$.

\begin{figure}[t]
\centering
\subfigure[]{
\includegraphics[width=.23\textwidth]{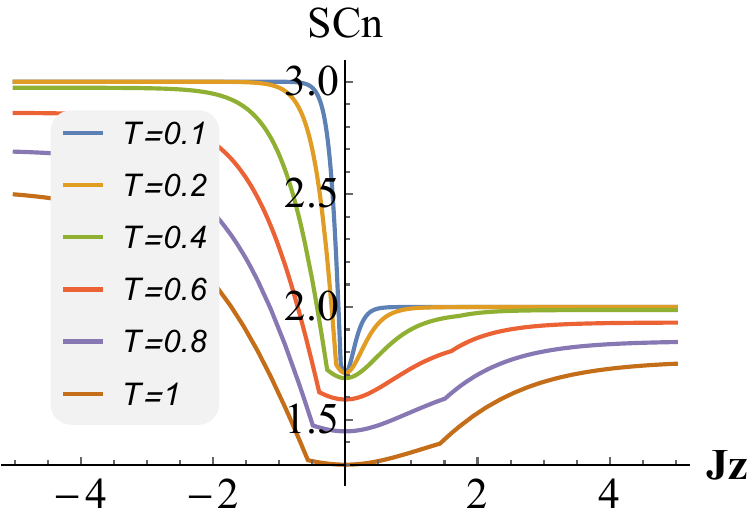}
}
\subfigure[]{
\includegraphics[width=.23\textwidth]{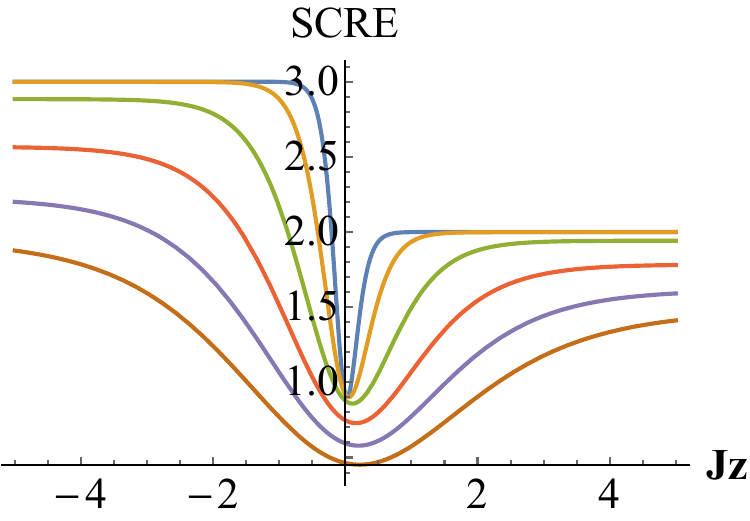}
}
\subfigure[]{
\includegraphics[width=.23\textwidth]{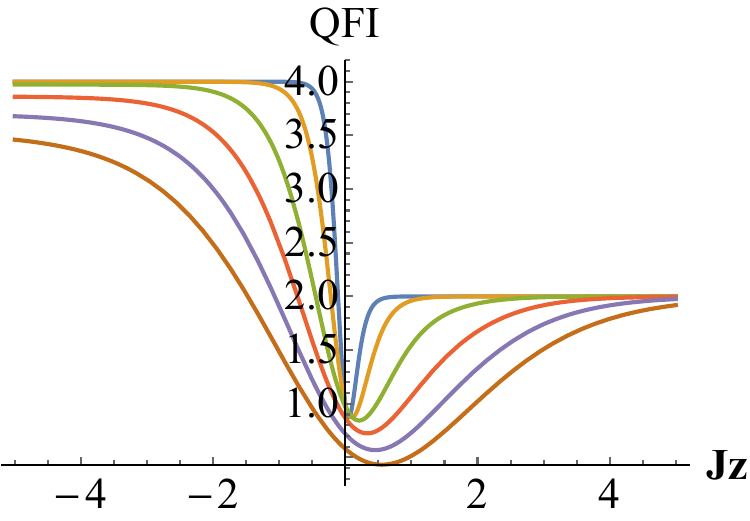}
}
\caption{The $l_1$ norm coherence of steered coherence, relative entropy
of steered coherence,
and quantum Fisher information vs the $J_z$ for
$B=1$, $J=1$ and $T=0.1, 0.2, 0.4, 0.6, 0.8,$ and $1$.}
\label{fig10}
\end{figure}

Fig. \ref{fig10} shows the variation of the three measures with respect to the interaction intensity $J_z$ at different temperatures $T=0.1, 0.2, 0.4, 0.6, 0.8, 1$, respectively, for fixed $B=1$ and $J=1$. One can see the different behavior of the three measures: Fig. \ref{fig10} (a) shows the symmetry around the vertical axes while Fig. \ref{fig10} (b) and (c) do not. With the increase of $J_z$, the SCn decreases to the minimum and then increases slowly to the maximum value of 2. At T=0.1, the SCRE gradually decreases from the maximum value of 3 and then slowly increases to 2, namely, its maximum value on the left-hand side is larger than the one on the right-hand side. QFI shows similar properties as SCRE with a maximum value of 4.

\section{Conclusions}
We have investigated three different quantum coherence measures, i.e., The $l_1$ norm coherence of steered coherence, relative entropy
of steered coherence, and the quantum Fisher information for spin chain systems.
The quantum behaviors in the $XXZ$ model have been demonstrated and analyzed.
The three measures characterize the quantum behaviors of the two-qubit Gibbs state
with respect to the variations of the temperature, external magnetic field, and interaction intensities.
The similarities among these behaviors have been demonstrated.
The results may highlight further investigations on related quantum measures and quantum spin systems.

\section*{Acknowledgments}
This work was supported in part by
the National Natural Science Foundation of China under Grant Nos. (12075159, 12171044, and 11905131), Beijing Natural Science Foundation (Z190005), the Academician
Innovation Platform of Hainan Province. This work was also supported by the Jiangxi Natural Science Foundation
20224BAB201027 and Shangrao City Science and Technology Plan Project 2020K009. We thank Prof. C.-P. Yang for helpful suggestions.


\begin{thebibliography}{42}%
\makeatletter
\providecommand \@ifxundefined [1]{%
 \@ifx{#1\undefined}
}%
\providecommand \@ifnum [1]{%
 \ifnum #1\expandafter \@firstoftwo
 \else \expandafter \@secondoftwo
 \fi
}%
\providecommand \@ifx [1]{%
 \ifx #1\expandafter \@firstoftwo
 \else \expandafter \@secondoftwo
 \fi
}%
\providecommand \natexlab [1]{#1}%
\providecommand \enquote  [1]{``#1''}%
\providecommand \bibnamefont  [1]{#1}%
\providecommand \bibfnamefont [1]{#1}%
\providecommand \citenamefont [1]{#1}%
\providecommand \href@noop [0]{\@secondoftwo}%
\providecommand \href [0]{\begingroup \@sanitize@url \@href}%
\providecommand \@href[1]{\@@startlink{#1}\@@href}%
\providecommand \@@href[1]{\endgroup#1\@@endlink}%
\providecommand \@sanitize@url [0]{\catcode `\\12\catcode `\$12\catcode
  `\&12\catcode `\#12\catcode `\^12\catcode `\_12\catcode `\%12\relax}%
\providecommand \@@startlink[1]{}%
\providecommand \@@endlink[0]{}%
\providecommand \url  [0]{\begingroup\@sanitize@url \@url }%
\providecommand \@url [1]{\endgroup\@href {#1}{\urlprefix }}%
\providecommand \urlprefix  [0]{URL }%
\providecommand \Eprint [0]{\href }%
\providecommand \doibase [0]{https://doi.org/}%
\providecommand \selectlanguage [0]{\@gobble}%
\providecommand \bibinfo  [0]{\@secondoftwo}%
\providecommand \bibfield  [0]{\@secondoftwo}%
\providecommand \translation [1]{[#1]}%
\providecommand \BibitemOpen [0]{}%
\providecommand \bibitemStop [0]{}%
\providecommand \bibitemNoStop [0]{.\EOS\space}%
\providecommand \EOS [0]{\spacefactor3000\relax}%
\providecommand \BibitemShut  [1]{\csname bibitem#1\endcsname}%
\let\auto@bib@innerbib\@empty
\bibitem [{\citenamefont {Sachdev}(1999)}]{Sachdev1999}%
  \BibitemOpen
  \bibfield  {author} {\bibinfo {author} {\bibfnamefont {S.}~\bibnamefont
  {Sachdev}},\ }\href@noop {} {\emph {\bibinfo {title} {Quantum Phase
  Transitions}}}\ (\bibinfo  {publisher} {Cambridge University Press ,
  Cambridge},\ \bibinfo {year} {1999})\BibitemShut {NoStop}%
\bibitem [{\citenamefont {Modi}\ \emph {et~al.}(2012)\citenamefont {Modi},
  \citenamefont {Brodutch}, \citenamefont {Cable}, \citenamefont {Paterek},\
  and\ \citenamefont {Vedral}}]{Modi2012}%
  \BibitemOpen
  \bibfield  {author} {\bibinfo {author} {\bibfnamefont {K.}~\bibnamefont
  {Modi}}, \bibinfo {author} {\bibfnamefont {A.}~\bibnamefont {Brodutch}},
  \bibinfo {author} {\bibfnamefont {H.}~\bibnamefont {Cable}}, \bibinfo
  {author} {\bibfnamefont {T.}~\bibnamefont {Paterek}},\ and\ \bibinfo {author}
  {\bibfnamefont {V.}~\bibnamefont {Vedral}},\ }\bibfield  {title} {\bibinfo
  {title} {The classical-quantum boundary for correlations: discord and related
  measures},\ }\href@noop {} {\bibfield  {journal} {\bibinfo  {journal} {Rev.
  Mod. Phys.}\ }\textbf {\bibinfo {volume} {84}},\ \bibinfo {pages} {1655}
  (\bibinfo {year} {2012})}\BibitemShut {NoStop}%
\bibitem [{\citenamefont {Adesso}\ \emph {et~al.}(2016)\citenamefont {Adesso},
  \citenamefont {Bromley},\ and\ \citenamefont {Cianciaruso}}]{Adesso2016}%
  \BibitemOpen
  \bibfield  {author} {\bibinfo {author} {\bibfnamefont {G.}~\bibnamefont
  {Adesso}}, \bibinfo {author} {\bibfnamefont {T.~R.}\ \bibnamefont
  {Bromley}},\ and\ \bibinfo {author} {\bibfnamefont {M.}~\bibnamefont
  {Cianciaruso}},\ }\bibfield  {title} {\bibinfo {title} {Measures and
  applications of quantum correlations},\ }\href@noop {} {\bibfield  {journal}
  {\bibinfo  {journal} {J. Phys. A: Math. Theor.}\ }\textbf {\bibinfo {volume}
  {49}},\ \bibinfo {pages} {473001} (\bibinfo {year} {2016})}\BibitemShut
  {NoStop}%
\bibitem [{\citenamefont {Nielsen}\ and\ \citenamefont
  {Chuang}(2010)}]{Nielsen2010}%
  \BibitemOpen
  \bibfield  {author} {\bibinfo {author} {\bibfnamefont {M.~A.}\ \bibnamefont
  {Nielsen}}\ and\ \bibinfo {author} {\bibfnamefont {I.~L.}\ \bibnamefont
  {Chuang}},\ }\href@noop {} {\emph {\bibinfo {title} {Quantum {C}omputation
  and {Q}uantum {I}nformation}}}\ (\bibinfo  {publisher} {Cambridge University
  Press},\ \bibinfo {year} {2010})\BibitemShut {NoStop}%
\bibitem [{\citenamefont {Ollivier}\ and\ \citenamefont
  {Zurek}(2001)}]{Ollivier2001}%
  \BibitemOpen
  \bibfield  {author} {\bibinfo {author} {\bibfnamefont {H.}~\bibnamefont
  {Ollivier}}\ and\ \bibinfo {author} {\bibfnamefont {W.~H.}\ \bibnamefont
  {Zurek}},\ }\bibfield  {title} {\bibinfo {title} {Quantum discord: a measure
  of the quantumness of correlations},\ }\href@noop {} {\bibfield  {journal}
  {\bibinfo  {journal} {Phys. Rev. Lett.}\ }\textbf {\bibinfo {volume} {88}},\
  \bibinfo {pages} {017901} (\bibinfo {year} {2001})}\BibitemShut {NoStop}%
\bibitem [{\citenamefont {Daki{\'c}}\ \emph {et~al.}(2010)\citenamefont
  {Daki{\'c}}, \citenamefont {Vedral},\ and\ \citenamefont
  {Brukner}}]{Dakic2010}%
  \BibitemOpen
  \bibfield  {author} {\bibinfo {author} {\bibfnamefont {B.}~\bibnamefont
  {Daki{\'c}}}, \bibinfo {author} {\bibfnamefont {V.}~\bibnamefont {Vedral}},\
  and\ \bibinfo {author} {\bibfnamefont {{\v{C}}.}~\bibnamefont {Brukner}},\
  }\bibfield  {title} {\bibinfo {title} {Necessary and sufficient condition for
  nonzero quantum discord},\ }\href@noop {} {\bibfield  {journal} {\bibinfo
  {journal} {Phys. Rev. Lett.}\ }\textbf {\bibinfo {volume} {105}},\ \bibinfo
  {pages} {190502} (\bibinfo {year} {2010})}\BibitemShut {NoStop}%
\bibitem [{\citenamefont {Luo}(2008)}]{Luo2008}%
  \BibitemOpen
  \bibfield  {author} {\bibinfo {author} {\bibfnamefont {S.}~\bibnamefont
  {Luo}},\ }\bibfield  {title} {\bibinfo {title} {Using measurement-induced
  disturbance to characterize correlations as classical or quantum},\
  }\href@noop {} {\bibfield  {journal} {\bibinfo  {journal} {Phys. Rev. A}\
  }\textbf {\bibinfo {volume} {77}},\ \bibinfo {pages} {022301} (\bibinfo
  {year} {2008})}\BibitemShut {NoStop}%
\bibitem [{\citenamefont {Rulli}\ and\ \citenamefont
  {Sarandy}(2011)}]{Rulli2011}%
  \BibitemOpen
  \bibfield  {author} {\bibinfo {author} {\bibfnamefont {C.~C.}\ \bibnamefont
  {Rulli}}\ and\ \bibinfo {author} {\bibfnamefont {M.~S.}\ \bibnamefont
  {Sarandy}},\ }\bibfield  {title} {\bibinfo {title} {Global quantum discord in
  multipartite systems},\ }\href {https://doi.org/10.1103/PhysRevA.84.042109}
  {\bibfield  {journal} {\bibinfo  {journal} {Phys. Rev. A}\ }\textbf {\bibinfo
  {volume} {84}},\ \bibinfo {pages} {042109} (\bibinfo {year}
  {2011})}\BibitemShut {NoStop}%
\bibitem [{\citenamefont {Bennett}\ \emph {et~al.}(1993)\citenamefont
  {Bennett}, \citenamefont {Brassard}, \citenamefont {Cr{\'e}peau},
  \citenamefont {Jozsa}, \citenamefont {Peres},\ and\ \citenamefont
  {Wootters}}]{Bennett1993}%
  \BibitemOpen
  \bibfield  {author} {\bibinfo {author} {\bibfnamefont {C.~H.}\ \bibnamefont
  {Bennett}}, \bibinfo {author} {\bibfnamefont {G.}~\bibnamefont {Brassard}},
  \bibinfo {author} {\bibfnamefont {C.}~\bibnamefont {Cr{\'e}peau}}, \bibinfo
  {author} {\bibfnamefont {R.}~\bibnamefont {Jozsa}}, \bibinfo {author}
  {\bibfnamefont {A.}~\bibnamefont {Peres}},\ and\ \bibinfo {author}
  {\bibfnamefont {W.~K.}\ \bibnamefont {Wootters}},\ }\bibfield  {title}
  {\bibinfo {title} {Teleporting an unknown quantum state via dual classical
  and {E}instein-{P}odolsky-{R}osen channels},\ }\href@noop {} {\bibfield
  {journal} {\bibinfo  {journal} {Phys. Rev. Lett.}\ }\textbf {\bibinfo
  {volume} {70}},\ \bibinfo {pages} {1895} (\bibinfo {year}
  {1993})}\BibitemShut {NoStop}%
\bibitem [{\citenamefont {Wang}\ \emph {et~al.}(2005)\citenamefont {Wang},
  \citenamefont {Deng}, \citenamefont {Li}, \citenamefont {Liu},\ and\
  \citenamefont {Long}}]{Wang2005}%
  \BibitemOpen
  \bibfield  {author} {\bibinfo {author} {\bibfnamefont {C.}~\bibnamefont
  {Wang}}, \bibinfo {author} {\bibfnamefont {F.-G.}\ \bibnamefont {Deng}},
  \bibinfo {author} {\bibfnamefont {Y.-S.}\ \bibnamefont {Li}}, \bibinfo
  {author} {\bibfnamefont {X.-S.}\ \bibnamefont {Liu}},\ and\ \bibinfo {author}
  {\bibfnamefont {G.~L.}\ \bibnamefont {Long}},\ }\bibfield  {title} {\bibinfo
  {title} {Quantum secure direct communication with high-dimension quantum
  superdense coding},\ }\href {https://doi.org/10.1103/PhysRevA.71.044305}
  {\bibfield  {journal} {\bibinfo  {journal} {Phys. Rev. A}\ }\textbf {\bibinfo
  {volume} {71}},\ \bibinfo {pages} {044305} (\bibinfo {year}
  {2005})}\BibitemShut {NoStop}%
\bibitem [{\citenamefont {Yu}\ \emph {et~al.}(2020)\citenamefont {Yu},
  \citenamefont {Ma}, \citenamefont {Luo}, \citenamefont {Jing}, \citenamefont
  {Sun}, \citenamefont {Fang}, \citenamefont {Yang}, \citenamefont {Liu},
  \citenamefont {Zheng}, \citenamefont {Xie} \emph {et~al.}}]{Yu2020}%
  \BibitemOpen
  \bibfield  {author} {\bibinfo {author} {\bibfnamefont {Y.}~\bibnamefont
  {Yu}}, \bibinfo {author} {\bibfnamefont {F.}~\bibnamefont {Ma}}, \bibinfo
  {author} {\bibfnamefont {X.-Y.}\ \bibnamefont {Luo}}, \bibinfo {author}
  {\bibfnamefont {B.}~\bibnamefont {Jing}}, \bibinfo {author} {\bibfnamefont
  {P.-F.}\ \bibnamefont {Sun}}, \bibinfo {author} {\bibfnamefont {R.-Z.}\
  \bibnamefont {Fang}}, \bibinfo {author} {\bibfnamefont {C.-W.}\ \bibnamefont
  {Yang}}, \bibinfo {author} {\bibfnamefont {H.}~\bibnamefont {Liu}}, \bibinfo
  {author} {\bibfnamefont {M.-Y.}\ \bibnamefont {Zheng}}, \bibinfo {author}
  {\bibfnamefont {X.-P.}\ \bibnamefont {Xie}}, \emph {et~al.},\ }\bibfield
  {title} {\bibinfo {title} {Entanglement of two quantum memories via fibres
  over dozens of kilometres},\ }\href@noop {} {\bibfield  {journal} {\bibinfo
  {journal} {Nature}\ }\textbf {\bibinfo {volume} {578}},\ \bibinfo {pages}
  {240} (\bibinfo {year} {2020})}\BibitemShut {NoStop}%
\bibitem [{\citenamefont {Daki{\'c}}\ \emph {et~al.}(2012)\citenamefont
  {Daki{\'c}}, \citenamefont {Lipp}, \citenamefont {Ma}, \citenamefont
  {Ringbauer}, \citenamefont {Kropatschek}, \citenamefont {Barz}, \citenamefont
  {Paterek}, \citenamefont {Vedral}, \citenamefont {Zeilinger}, \citenamefont
  {Brukner} \emph {et~al.}}]{Dakic2012}%
  \BibitemOpen
  \bibfield  {author} {\bibinfo {author} {\bibfnamefont {B.}~\bibnamefont
  {Daki{\'c}}}, \bibinfo {author} {\bibfnamefont {Y.~O.}\ \bibnamefont {Lipp}},
  \bibinfo {author} {\bibfnamefont {X.}~\bibnamefont {Ma}}, \bibinfo {author}
  {\bibfnamefont {M.}~\bibnamefont {Ringbauer}}, \bibinfo {author}
  {\bibfnamefont {S.}~\bibnamefont {Kropatschek}}, \bibinfo {author}
  {\bibfnamefont {S.}~\bibnamefont {Barz}}, \bibinfo {author} {\bibfnamefont
  {T.}~\bibnamefont {Paterek}}, \bibinfo {author} {\bibfnamefont
  {V.}~\bibnamefont {Vedral}}, \bibinfo {author} {\bibfnamefont
  {A.}~\bibnamefont {Zeilinger}}, \bibinfo {author} {\bibfnamefont
  {{\v{C}}.}~\bibnamefont {Brukner}}, \emph {et~al.},\ }\bibfield  {title}
  {\bibinfo {title} {Quantum discord as resource for remote state
  preparation},\ }\href@noop {} {\bibfield  {journal} {\bibinfo  {journal}
  {Nat. Phys.}\ }\textbf {\bibinfo {volume} {8}},\ \bibinfo {pages} {666}
  (\bibinfo {year} {2012})}\BibitemShut {NoStop}%
\bibitem [{\citenamefont {Li}\ \emph {et~al.}(2012)\citenamefont {Li},
  \citenamefont {Fei}, \citenamefont {Wang},\ and\ \citenamefont
  {Fan}}]{Li2012}%
  \BibitemOpen
  \bibfield  {author} {\bibinfo {author} {\bibfnamefont {B.}~\bibnamefont
  {Li}}, \bibinfo {author} {\bibfnamefont {S.-M.}\ \bibnamefont {Fei}},
  \bibinfo {author} {\bibfnamefont {Z.-X.}\ \bibnamefont {Wang}},\ and\
  \bibinfo {author} {\bibfnamefont {H.}~\bibnamefont {Fan}},\ }\bibfield
  {title} {\bibinfo {title} {Assisted state discrimination without
  entanglement},\ }\href {https://doi.org/10.1103/PhysRevA.85.022328}
  {\bibfield  {journal} {\bibinfo  {journal} {Phys. Rev. A}\ }\textbf {\bibinfo
  {volume} {85}},\ \bibinfo {pages} {022328} (\bibinfo {year}
  {2012})}\BibitemShut {NoStop}%
\bibitem [{\citenamefont {Madhok}\ and\ \citenamefont
  {Datta}(2011)}]{Madhok2011}%
  \BibitemOpen
  \bibfield  {author} {\bibinfo {author} {\bibfnamefont {V.}~\bibnamefont
  {Madhok}}\ and\ \bibinfo {author} {\bibfnamefont {A.}~\bibnamefont {Datta}},\
  }\bibfield  {title} {\bibinfo {title} {Interpreting quantum discord through
  quantum state merging},\ }\href {https://doi.org/10.1103/PhysRevA.83.032323}
  {\bibfield  {journal} {\bibinfo  {journal} {Phys. Rev. A}\ }\textbf {\bibinfo
  {volume} {83}},\ \bibinfo {pages} {032323} (\bibinfo {year}
  {2011})}\BibitemShut {NoStop}%
\bibitem [{\citenamefont {{\c{C}}akmak}\ \emph {et~al.}(2015)\citenamefont
  {{\c{C}}akmak}, \citenamefont {Karpat},\ and\ \citenamefont
  {Fanchini}}]{Cakmak2015}%
  \BibitemOpen
  \bibfield  {author} {\bibinfo {author} {\bibfnamefont {B.}~\bibnamefont
  {{\c{C}}akmak}}, \bibinfo {author} {\bibfnamefont {G.}~\bibnamefont
  {Karpat}},\ and\ \bibinfo {author} {\bibfnamefont {F.~F.}\ \bibnamefont
  {Fanchini}},\ }\bibfield  {title} {\bibinfo {title} {Factorization and
  criticality in the anisotropic {XY} chain via correlations},\ }\href@noop {}
  {\bibfield  {journal} {\bibinfo  {journal} {Entropy}\ }\textbf {\bibinfo
  {volume} {17}},\ \bibinfo {pages} {790} (\bibinfo {year} {2015})}\BibitemShut
  {NoStop}%
\bibitem [{\citenamefont {Barouch}\ \emph {et~al.}(1970)\citenamefont
  {Barouch}, \citenamefont {McCoy},\ and\ \citenamefont
  {Dresden}}]{Barouch1970}%
  \BibitemOpen
  \bibfield  {author} {\bibinfo {author} {\bibfnamefont {E.}~\bibnamefont
  {Barouch}}, \bibinfo {author} {\bibfnamefont {B.~M.}\ \bibnamefont {McCoy}},\
  and\ \bibinfo {author} {\bibfnamefont {M.}~\bibnamefont {Dresden}},\
  }\bibfield  {title} {\bibinfo {title} {Statistical mechanics of the {XY}
  model. {I}},\ }\href@noop {} {\bibfield  {journal} {\bibinfo  {journal}
  {Phys. Rev. A}\ }\textbf {\bibinfo {volume} {2}},\ \bibinfo {pages} {1075}
  (\bibinfo {year} {1970})}\BibitemShut {NoStop}%
\bibitem [{\citenamefont {Barouch}\ and\ \citenamefont
  {McCoy}(1971)}]{Barouch1971}%
  \BibitemOpen
  \bibfield  {author} {\bibinfo {author} {\bibfnamefont {E.}~\bibnamefont
  {Barouch}}\ and\ \bibinfo {author} {\bibfnamefont {B.~M.}\ \bibnamefont
  {McCoy}},\ }\bibfield  {title} {\bibinfo {title} {Statistical mechanics of
  the {XY} model. {II}. spin-correlation functions},\ }\href@noop {} {\bibfield
   {journal} {\bibinfo  {journal} {Phys. Rev. A}\ }\textbf {\bibinfo {volume}
  {3}},\ \bibinfo {pages} {786} (\bibinfo {year} {1971})}\BibitemShut {NoStop}%
\bibitem [{\citenamefont {Yurischev}(2020)}]{AYurischev2020}%
  \BibitemOpen
  \bibfield  {author} {\bibinfo {author} {\bibfnamefont {M.}~\bibnamefont
  {Yurischev}},\ }\bibfield  {title} {\bibinfo {title} {{Temperature-field
  phase diagrams of one-way quantum work deficit in two-qubit {XXZ} spin
  systems}},\ }\href {https://doi.org/10.1007/s11128-020-2610-1} {\bibfield
  {journal} {\bibinfo  {journal} {Quantum Inf. Process.}\ }\textbf {\bibinfo
  {volume} {19}},\ \bibinfo {pages} {110} (\bibinfo {year} {2020})}\BibitemShut
  {NoStop}%
\bibitem [{\citenamefont {Arnesen}\ \emph {et~al.}(2001)\citenamefont
  {Arnesen}, \citenamefont {Bose},\ and\ \citenamefont {Vedral}}]{Arnesen2001}%
  \BibitemOpen
  \bibfield  {author} {\bibinfo {author} {\bibfnamefont {M.}~\bibnamefont
  {Arnesen}}, \bibinfo {author} {\bibfnamefont {S.}~\bibnamefont {Bose}},\ and\
  \bibinfo {author} {\bibfnamefont {V.}~\bibnamefont {Vedral}},\ }\bibfield
  {title} {\bibinfo {title} {Natural thermal and magnetic entanglement in the
  1{D} {H}eisenberg model},\ }\href@noop {} {\bibfield  {journal} {\bibinfo
  {journal} {Phys. Rev. Lett.}\ }\textbf {\bibinfo {volume} {87}},\ \bibinfo
  {pages} {017901} (\bibinfo {year} {2001})}\BibitemShut {NoStop}%
\bibitem [{\citenamefont {Maziero}\ \emph {et~al.}(2010)\citenamefont
  {Maziero}, \citenamefont {Guzman}, \citenamefont {C{\'e}leri}, \citenamefont
  {Sarandy},\ and\ \citenamefont {Serra}}]{Maziero2010}%
  \BibitemOpen
  \bibfield  {author} {\bibinfo {author} {\bibfnamefont {J.}~\bibnamefont
  {Maziero}}, \bibinfo {author} {\bibfnamefont {H.}~\bibnamefont {Guzman}},
  \bibinfo {author} {\bibfnamefont {L.}~\bibnamefont {C{\'e}leri}}, \bibinfo
  {author} {\bibfnamefont {M.}~\bibnamefont {Sarandy}},\ and\ \bibinfo {author}
  {\bibfnamefont {R.}~\bibnamefont {Serra}},\ }\bibfield  {title} {\bibinfo
  {title} {Quantum and classical thermal correlations in the {XY} spin-1/2
  chain},\ }\href@noop {} {\bibfield  {journal} {\bibinfo  {journal} {Phys.
  Rev. A}\ }\textbf {\bibinfo {volume} {82}},\ \bibinfo {pages} {012106}
  (\bibinfo {year} {2010})}\BibitemShut {NoStop}%
\bibitem [{\citenamefont {Ye}\ \emph {et~al.}(2017)\citenamefont {Ye},
  \citenamefont {Li}, \citenamefont {Zhao}, \citenamefont {Zhang},\ and\
  \citenamefont {Fei}}]{Ye2017}%
  \BibitemOpen
  \bibfield  {author} {\bibinfo {author} {\bibfnamefont {B.-L.}\ \bibnamefont
  {Ye}}, \bibinfo {author} {\bibfnamefont {B.}~\bibnamefont {Li}}, \bibinfo
  {author} {\bibfnamefont {L.-J.}\ \bibnamefont {Zhao}}, \bibinfo {author}
  {\bibfnamefont {H.-J.}\ \bibnamefont {Zhang}},\ and\ \bibinfo {author}
  {\bibfnamefont {S.-M.}\ \bibnamefont {Fei}},\ }\bibfield  {title} {\bibinfo
  {title} {One-way quantum deficit and quantum coherence in the anisotropic
  ${XY}$ chain},\ }\href@noop {} {\bibfield  {journal} {\bibinfo  {journal}
  {Sci. China-Phys. Mech. Astron.}\ }\textbf {\bibinfo {volume} {60}},\
  \bibinfo {pages} {030311} (\bibinfo {year} {2017})}\BibitemShut {NoStop}%
\bibitem [{\citenamefont {Shan}\ \emph {et~al.}(2014)\citenamefont {Shan},
  \citenamefont {Cheng}, \citenamefont {Liu}, \citenamefont {Cheng},\ and\
  \citenamefont {Liu}}]{Shan2014}%
  \BibitemOpen
  \bibfield  {author} {\bibinfo {author} {\bibfnamefont {C.-J.}\ \bibnamefont
  {Shan}}, \bibinfo {author} {\bibfnamefont {W.-W.}\ \bibnamefont {Cheng}},
  \bibinfo {author} {\bibfnamefont {J.-B.}\ \bibnamefont {Liu}}, \bibinfo
  {author} {\bibfnamefont {Y.-S.}\ \bibnamefont {Cheng}},\ and\ \bibinfo
  {author} {\bibfnamefont {T.-K.}\ \bibnamefont {Liu}},\ }\bibfield  {title}
  {\bibinfo {title} {Scaling of geometric quantum discord close to a
  topological phase transition},\ }\href@noop {} {\bibfield  {journal}
  {\bibinfo  {journal} {Sci. Rep.}\ }\textbf {\bibinfo {volume} {4}},\ \bibinfo
  {pages} {1} (\bibinfo {year} {2014})}\BibitemShut {NoStop}%
\bibitem [{\citenamefont {Altintas}\ and\ \citenamefont
  {Eryigit}(2012)}]{Altintas2012}%
  \BibitemOpen
  \bibfield  {author} {\bibinfo {author} {\bibfnamefont {F.}~\bibnamefont
  {Altintas}}\ and\ \bibinfo {author} {\bibfnamefont {R.}~\bibnamefont
  {Eryigit}},\ }\bibfield  {title} {\bibinfo {title} {Correlation and
  nonlocality measures as indicators of quantum phase transitions in several
  critical systems},\ }\href@noop {} {\bibfield  {journal} {\bibinfo  {journal}
  {Ann. Phys.}\ }\textbf {\bibinfo {volume} {327}},\ \bibinfo {pages} {3084}
  (\bibinfo {year} {2012})}\BibitemShut {NoStop}%
\bibitem [{\citenamefont {Sha}\ \emph {et~al.}(2018)\citenamefont {Sha},
  \citenamefont {Wang}, \citenamefont {Sun},\ and\ \citenamefont
  {Hou}}]{Sha2018}%
  \BibitemOpen
  \bibfield  {author} {\bibinfo {author} {\bibfnamefont {Y.-T.}\ \bibnamefont
  {Sha}}, \bibinfo {author} {\bibfnamefont {Y.}~\bibnamefont {Wang}}, \bibinfo
  {author} {\bibfnamefont {Z.-H.}\ \bibnamefont {Sun}},\ and\ \bibinfo {author}
  {\bibfnamefont {X.-W.}\ \bibnamefont {Hou}},\ }\bibfield  {title} {\bibinfo
  {title} {Thermal quantum coherence and correlation in the extended {$XY$}
  spin chain},\ }\href@noop {} {\bibfield  {journal} {\bibinfo  {journal} {Ann.
  Phys.}\ }\textbf {\bibinfo {volume} {392}},\ \bibinfo {pages} {229} (\bibinfo
  {year} {2018})}\BibitemShut {NoStop}%
\bibitem [{\citenamefont {Li}\ and\ \citenamefont {Lin}(2016)}]{li2016quantum}%
  \BibitemOpen
  \bibfield  {author} {\bibinfo {author} {\bibfnamefont {Y.-C.}\ \bibnamefont
  {Li}}\ and\ \bibinfo {author} {\bibfnamefont {H.-Q.}\ \bibnamefont {Lin}},\
  }\bibfield  {title} {\bibinfo {title} {Quantum coherence and quantum phase
  transitions},\ }\href@noop {} {\bibfield  {journal} {\bibinfo  {journal}
  {Sci. Rep.}\ }\textbf {\bibinfo {volume} {6}},\ \bibinfo {pages} {26365}
  (\bibinfo {year} {2016})}\BibitemShut {NoStop}%
\bibitem [{\citenamefont {Yin}\ \emph {et~al.}(2018)\citenamefont {Yin},
  \citenamefont {Song}, \citenamefont {Xu}, \citenamefont {Zhang},\ and\
  \citenamefont {Liu}}]{yin2018quantum}%
  \BibitemOpen
  \bibfield  {author} {\bibinfo {author} {\bibfnamefont {S.}~\bibnamefont
  {Yin}}, \bibinfo {author} {\bibfnamefont {J.}~\bibnamefont {Song}}, \bibinfo
  {author} {\bibfnamefont {X.}~\bibnamefont {Xu}}, \bibinfo {author}
  {\bibfnamefont {Y.}~\bibnamefont {Zhang}},\ and\ \bibinfo {author}
  {\bibfnamefont {S.}~\bibnamefont {Liu}},\ }\bibfield  {title} {\bibinfo
  {title} {Quantum coherence dynamics of three-qubit states in {$XY$}
  spin-chain environment},\ }\href@noop {} {\bibfield  {journal} {\bibinfo
  {journal} {Quantum Inf. Process.}\ }\textbf {\bibinfo {volume} {17}},\
  \bibinfo {pages} {1} (\bibinfo {year} {2018})}\BibitemShut {NoStop}%
\bibitem [{\citenamefont {Zhang}\ and\ \citenamefont
  {Xu}(2017)}]{zhang2017quantum}%
  \BibitemOpen
  \bibfield  {author} {\bibinfo {author} {\bibfnamefont {G.-Q.}\ \bibnamefont
  {Zhang}}\ and\ \bibinfo {author} {\bibfnamefont {J.-B.}\ \bibnamefont {Xu}},\
  }\bibfield  {title} {\bibinfo {title} {Quantum coherence of an {$XY$} spin
  chain with {D}zyaloshinskii-{M}oriya interaction and quantum phase
  transition},\ }\href@noop {} {\bibfield  {journal} {\bibinfo  {journal} {J.
  Phys. A: Math. Theor.}\ }\textbf {\bibinfo {volume} {50}},\ \bibinfo {pages}
  {265303} (\bibinfo {year} {2017})}\BibitemShut {NoStop}%
\bibitem [{\citenamefont {Jafari}\ and\ \citenamefont
  {Akbari}(2020)}]{jafari2020dynamics}%
  \BibitemOpen
  \bibfield  {author} {\bibinfo {author} {\bibfnamefont {R.}~\bibnamefont
  {Jafari}}\ and\ \bibinfo {author} {\bibfnamefont {A.}~\bibnamefont
  {Akbari}},\ }\bibfield  {title} {\bibinfo {title} {Dynamics of quantum
  coherence and quantum {F}isher information after a sudden quench},\
  }\href@noop {} {\bibfield  {journal} {\bibinfo  {journal} {Phys. Rev. A}\
  }\textbf {\bibinfo {volume} {101}},\ \bibinfo {pages} {062105} (\bibinfo
  {year} {2020})}\BibitemShut {NoStop}%
\bibitem [{\citenamefont {Qin}\ \emph {et~al.}(2018)\citenamefont {Qin},
  \citenamefont {Ren},\ and\ \citenamefont {Zhang}}]{qin2018dynamics}%
  \BibitemOpen
  \bibfield  {author} {\bibinfo {author} {\bibfnamefont {M.}~\bibnamefont
  {Qin}}, \bibinfo {author} {\bibfnamefont {Z.}~\bibnamefont {Ren}},\ and\
  \bibinfo {author} {\bibfnamefont {X.}~\bibnamefont {Zhang}},\ }\bibfield
  {title} {\bibinfo {title} {Dynamics of quantum coherence and quantum phase
  transitions in {XY} spin systems},\ }\href@noop {} {\bibfield  {journal}
  {\bibinfo  {journal} {Phys. Rev. A}\ }\textbf {\bibinfo {volume} {98}},\
  \bibinfo {pages} {012303} (\bibinfo {year} {2018})}\BibitemShut {NoStop}%
\bibitem [{\citenamefont {Lu}\ and\ \citenamefont {Wang}(2021)}]{Lu2021}%
  \BibitemOpen
  \bibfield  {author} {\bibinfo {author} {\bibfnamefont {X.-M.}\ \bibnamefont
  {Lu}}\ and\ \bibinfo {author} {\bibfnamefont {X.}~\bibnamefont {Wang}},\
  }\bibfield  {title} {\bibinfo {title} {Incorporating {H}eisenberg's
  uncertainty principle into quantum multiparameter estimation},\ }\href
  {https://doi.org/10.1103/PhysRevLett.126.120503} {\bibfield  {journal}
  {\bibinfo  {journal} {Phys. Rev. Lett.}\ }\textbf {\bibinfo {volume} {126}},\
  \bibinfo {pages} {120503} (\bibinfo {year} {2021})}\BibitemShut {NoStop}%
\bibitem [{\citenamefont {Hu}\ \emph {et~al.}(2021)\citenamefont {Hu},
  \citenamefont {Fang},\ and\ \citenamefont {Fan}}]{Hu2021}%
  \BibitemOpen
  \bibfield  {author} {\bibinfo {author} {\bibfnamefont {M.~L.}\ \bibnamefont
  {Hu}}, \bibinfo {author} {\bibfnamefont {F.}~\bibnamefont {Fang}},\ and\
  \bibinfo {author} {\bibfnamefont {H.}~\bibnamefont {Fan}},\ }\bibfield
  {title} {\bibinfo {title} {{Finite-size scaling of coherence and steered
  coherence in the Lipkin-Meshkov-Glick model}},\ }\href
  {https://doi.org/10.1103/PhysRevA.104.062416} {\bibfield  {journal} {\bibinfo
   {journal} {Phys. Rev. A}\ }\textbf {\bibinfo {volume} {104}},\ \bibinfo
  {pages} {062416} (\bibinfo {year} {2021})}\BibitemShut {NoStop}%
\bibitem [{\citenamefont {Zhao}\ \emph {et~al.}(2022)\citenamefont {Zhao},
  \citenamefont {Yi}, \citenamefont {Xue},\ and\ \citenamefont
  {You}}]{Zhao2022}%
  \BibitemOpen
  \bibfield  {author} {\bibinfo {author} {\bibfnamefont {Z.}~\bibnamefont
  {Zhao}}, \bibinfo {author} {\bibfnamefont {T.~C.}\ \bibnamefont {Yi}},
  \bibinfo {author} {\bibfnamefont {M.}~\bibnamefont {Xue}},\ and\ \bibinfo
  {author} {\bibfnamefont {W.~L.}\ \bibnamefont {You}},\ }\bibfield  {title}
  {\bibinfo {title} {{Characterizing quantum criticality and steered coherence
  in the XY-Gamma chain}},\ }\href
  {https://doi.org/10.1103/PhysRevA.105.063306} {\bibfield  {journal} {\bibinfo
   {journal} {Phys. Rev. A}\ }\textbf {\bibinfo {volume} {105}},\ \bibinfo
  {pages} {063306} (\bibinfo {year} {2022})}\BibitemShut {NoStop}%
\bibitem [{\citenamefont {Liu}\ and\ \citenamefont {Hu}(2023)}]{Liu2023}%
  \BibitemOpen
  \bibfield  {author} {\bibinfo {author} {\bibfnamefont {X.-Y.}\ \bibnamefont
  {Liu}}\ and\ \bibinfo {author} {\bibfnamefont {M.-L.}\ \bibnamefont {Hu}},\
  }\bibfield  {title} {\bibinfo {title} {{Average quantum coherence and its use
  in probing quantum phase transitions}},\ }\href
  {https://doi.org/10.1016/j.physa.2022.128308} {\bibfield  {journal} {\bibinfo
   {journal} {Physica A}\ }\textbf {\bibinfo {volume} {609}},\ \bibinfo {pages}
  {128308} (\bibinfo {year} {2023})}\BibitemShut {NoStop}%
\bibitem [{\citenamefont {Streltsov}\ \emph {et~al.}(2017)\citenamefont
  {Streltsov}, \citenamefont {Adesso},\ and\ \citenamefont
  {Plenio}}]{Streltsov2017}%
  \BibitemOpen
  \bibfield  {author} {\bibinfo {author} {\bibfnamefont {A.}~\bibnamefont
  {Streltsov}}, \bibinfo {author} {\bibfnamefont {G.}~\bibnamefont {Adesso}},\
  and\ \bibinfo {author} {\bibfnamefont {M.~B.}\ \bibnamefont {Plenio}},\
  }\bibfield  {title} {\bibinfo {title} {{Colloquium: Quantum coherence as a
  resource}},\ }\href {https://doi.org/10.1103/RevModPhys.89.041003} {\bibfield
   {journal} {\bibinfo  {journal} {Rev. Mod. Phys.}\ }\textbf {\bibinfo
  {volume} {89}},\ \bibinfo {pages} {041003} (\bibinfo {year}
  {2017})}\BibitemShut {NoStop}%
\bibitem [{\citenamefont {Zheng}\ \emph {et~al.}(2018)\citenamefont {Zheng},
  \citenamefont {Ma}, \citenamefont {Wang}, \citenamefont {Fei},\ and\
  \citenamefont {Peng}}]{Zheng2018}%
  \BibitemOpen
  \bibfield  {author} {\bibinfo {author} {\bibfnamefont {W.}~\bibnamefont
  {Zheng}}, \bibinfo {author} {\bibfnamefont {Z.}~\bibnamefont {Ma}}, \bibinfo
  {author} {\bibfnamefont {H.}~\bibnamefont {Wang}}, \bibinfo {author}
  {\bibfnamefont {S.-M.}\ \bibnamefont {Fei}},\ and\ \bibinfo {author}
  {\bibfnamefont {X.}~\bibnamefont {Peng}},\ }\bibfield  {title} {\bibinfo
  {title} {Experimental demonstration of observability and operability of
  robustness of coherence},\ }\href@noop {} {\bibfield  {journal} {\bibinfo
  {journal} {Phys. Rev. Lett.}\ }\textbf {\bibinfo {volume} {120}},\ \bibinfo
  {pages} {230504} (\bibinfo {year} {2018})}\BibitemShut {NoStop}%
\bibitem [{\citenamefont {Smirne}\ \emph {et~al.}(2020)\citenamefont {Smirne},
  \citenamefont {Nitsche}, \citenamefont {Egloff}, \citenamefont {Barkhofen},
  \citenamefont {De}, \citenamefont {Dhand}, \citenamefont {Silberhorn},
  \citenamefont {Huelga},\ and\ \citenamefont {Plenio}}]{Smirne2020}%
  \BibitemOpen
  \bibfield  {author} {\bibinfo {author} {\bibfnamefont {A.}~\bibnamefont
  {Smirne}}, \bibinfo {author} {\bibfnamefont {T.}~\bibnamefont {Nitsche}},
  \bibinfo {author} {\bibfnamefont {D.}~\bibnamefont {Egloff}}, \bibinfo
  {author} {\bibfnamefont {S.}~\bibnamefont {Barkhofen}}, \bibinfo {author}
  {\bibfnamefont {S.}~\bibnamefont {De}}, \bibinfo {author} {\bibfnamefont
  {I.}~\bibnamefont {Dhand}}, \bibinfo {author} {\bibfnamefont
  {C.}~\bibnamefont {Silberhorn}}, \bibinfo {author} {\bibfnamefont {S.~F.}\
  \bibnamefont {Huelga}},\ and\ \bibinfo {author} {\bibfnamefont {M.~B.}\
  \bibnamefont {Plenio}},\ }\bibfield  {title} {\bibinfo {title} {Experimental
  control of the degree of non-classicality via quantum coherence},\
  }\href@noop {} {\bibfield  {journal} {\bibinfo  {journal} {Quantum Sci.
  Technol.}\ }\textbf {\bibinfo {volume} {5}},\ \bibinfo {pages} {04LT01}
  (\bibinfo {year} {2020})}\BibitemShut {NoStop}%
\bibitem [{\citenamefont {Baumgratz}\ \emph {et~al.}(2014)\citenamefont
  {Baumgratz}, \citenamefont {Cramer},\ and\ \citenamefont
  {Plenio}}]{Baumgratz2014}%
  \BibitemOpen
  \bibfield  {author} {\bibinfo {author} {\bibfnamefont {T.}~\bibnamefont
  {Baumgratz}}, \bibinfo {author} {\bibfnamefont {M.}~\bibnamefont {Cramer}},\
  and\ \bibinfo {author} {\bibfnamefont {M.}~\bibnamefont {Plenio}},\
  }\bibfield  {title} {\bibinfo {title} {Quantifying coherence},\ }\href@noop
  {} {\bibfield  {journal} {\bibinfo  {journal} {Phys. Rev. Lett.}\ }\textbf
  {\bibinfo {volume} {113}},\ \bibinfo {pages} {140401} (\bibinfo {year}
  {2014})}\BibitemShut {NoStop}%
\bibitem [{\citenamefont {Karpat}\ \emph {et~al.}(2014)\citenamefont {Karpat},
  \citenamefont {{\c{C}}akmak},\ and\ \citenamefont {Fanchini}}]{Karpat2014}%
  \BibitemOpen
  \bibfield  {author} {\bibinfo {author} {\bibfnamefont {G.}~\bibnamefont
  {Karpat}}, \bibinfo {author} {\bibfnamefont {B.}~\bibnamefont
  {{\c{C}}akmak}},\ and\ \bibinfo {author} {\bibfnamefont {F.}~\bibnamefont
  {Fanchini}},\ }\bibfield  {title} {\bibinfo {title} {Quantum coherence and
  uncertainty in the anisotropic {XY} chain},\ }\href@noop {} {\bibfield
  {journal} {\bibinfo  {journal} {Phys. Rev. B}\ }\textbf {\bibinfo {volume}
  {90}},\ \bibinfo {pages} {104431} (\bibinfo {year} {2014})}\BibitemShut
  {NoStop}%
\bibitem [{\citenamefont {Li}\ \emph {et~al.}(2020)\citenamefont {Li},
  \citenamefont {Zhang},\ and\ \citenamefont {Lin}}]{Li2020}%
  \BibitemOpen
  \bibfield  {author} {\bibinfo {author} {\bibfnamefont {Y.~C.}\ \bibnamefont
  {Li}}, \bibinfo {author} {\bibfnamefont {J.}~\bibnamefont {Zhang}},\ and\
  \bibinfo {author} {\bibfnamefont {H.-Q.}\ \bibnamefont {Lin}},\ }\bibfield
  {title} {\bibinfo {title} {{Quantum coherence spectrum and quantum phase
  transitions}},\ }\href {https://doi.org/10.1103/physrevb.101.115142}
  {\bibfield  {journal} {\bibinfo  {journal} {Phys. Rev. B}\ }\textbf {\bibinfo
  {volume} {101}},\ \bibinfo {pages} {115142} (\bibinfo {year}
  {2020})}\BibitemShut {NoStop}%
\bibitem [{\citenamefont {Hu}\ \emph {et~al.}(2020)\citenamefont {Hu},
  \citenamefont {Gao},\ and\ \citenamefont {Fan}}]{Hu2020}%
  \BibitemOpen
  \bibfield  {author} {\bibinfo {author} {\bibfnamefont {M.-L.}\ \bibnamefont
  {Hu}}, \bibinfo {author} {\bibfnamefont {Y.-Y.}\ \bibnamefont {Gao}},\ and\
  \bibinfo {author} {\bibfnamefont {H.}~\bibnamefont {Fan}},\ }\bibfield
  {title} {\bibinfo {title} {{Steered quantum coherence as a signature of
  quantum phase transitions in spin chains}},\ }\href
  {https://doi.org/10.1103/PhysRevA.101.032305} {\bibfield  {journal} {\bibinfo
   {journal} {Phys. Rev. A}\ }\textbf {\bibinfo {volume} {101}},\ \bibinfo
  {pages} {032305} (\bibinfo {year} {2020})}\BibitemShut {NoStop}%
\bibitem [{\citenamefont {Mondal}\ \emph {et~al.}(2017)\citenamefont {Mondal},
  \citenamefont {Pramanik},\ and\ \citenamefont {Pati}}]{Mondal2017}%
  \BibitemOpen
  \bibfield  {author} {\bibinfo {author} {\bibfnamefont {D.}~\bibnamefont
  {Mondal}}, \bibinfo {author} {\bibfnamefont {T.}~\bibnamefont {Pramanik}},\
  and\ \bibinfo {author} {\bibfnamefont {A.~K.}\ \bibnamefont {Pati}},\
  }\bibfield  {title} {\bibinfo {title} {Nonlocal advantage of quantum
  coherence},\ }\href {https://doi.org/10.1103/PhysRevA.95.010301} {\bibfield
  {journal} {\bibinfo  {journal} {Phys. Rev. A}\ }\textbf {\bibinfo {volume}
  {95}},\ \bibinfo {pages} {010301} (\bibinfo {year} {2017})}\BibitemShut
  {NoStop}%
\bibitem [{\citenamefont {Ye}\ \emph {et~al.}(2018)\citenamefont {Ye},
  \citenamefont {Li}, \citenamefont {Wang}, \citenamefont {Li-Jost},\ and\
  \citenamefont {Fei}}]{Ye2018b}%
  \BibitemOpen
  \bibfield  {author} {\bibinfo {author} {\bibfnamefont {B.-L.}\ \bibnamefont
  {Ye}}, \bibinfo {author} {\bibfnamefont {B.}~\bibnamefont {Li}}, \bibinfo
  {author} {\bibfnamefont {Z.-X.}\ \bibnamefont {Wang}}, \bibinfo {author}
  {\bibfnamefont {X.}~\bibnamefont {Li-Jost}},\ and\ \bibinfo {author}
  {\bibfnamefont {S.-M.}\ \bibnamefont {Fei}},\ }\bibfield  {title} {\bibinfo
  {title} {Quantum {F}isher information and coherence in one-dimensional ${XY}$
  spin models with {D}zyaloshinsky-{M}oriya interactions},\ }\href@noop {}
  {\bibfield  {journal} {\bibinfo  {journal} {Sci. China-Phys. Mech. Astron.}\
  }\textbf {\bibinfo {volume} {61}},\ \bibinfo {pages} {110312} (\bibinfo
  {year} {2018})}\BibitemShut {NoStop}%
\end{thebibliography}

%

\end{document}